\documentclass[
 aip,
 jcp,
 amsmath,amssymb,
 reprint,
 floatfix,
 superscriptaddress
]{revtex4-2}

\usepackage[utf8]{inputenc}
\usepackage[T1]{fontenc}
\usepackage{etoolbox}
\usepackage{booktabs}
\usepackage{graphicx}
\usepackage{microtype}
\usepackage{capt-of}
\usepackage{xspace}
\usepackage{dcolumn}
\usepackage{bm}
\usepackage{array}
\usepackage{tabularx}
\usepackage{longtable}
\usepackage{xcolor}
\usepackage{enumitem}
\usepackage{float}
\usepackage{bibunits}
\defaultbibliographystyle{aipnum4-2}
\defaultbibliography{master-control,ref}
\usepackage{hyperref}
\hypersetup{hidelinks,hypertexnames=false}

\makeatletter
\patchcmd{\endbibunit}{\@input{bu.aux}}{}{}{}
\let\auto@bib\@empty
\newcommand{\masterbibcontrols}{%
 \immediate\write\@bibunitaux{\string\citation{REVTEX42Control}}%
 \immediate\write\@bibunitaux{\string\citation{aip41Control}}%
}
\makeatother

\DeclareMathVersion{siptm}
\SetSymbolFont{operators}{siptm}{OT1}{ztmcm}{m}{n}
\SetSymbolFont{letters}{siptm}{OML}{ztmcm}{m}{it}
\SetSymbolFont{symbols}{siptm}{OMS}{ztmcm}{m}{n}
\SetSymbolFont{largesymbols}{siptm}{OMX}{ztmcm}{m}{n}
\SetMathAlphabet{\mathbf}{siptm}{OT1}{ptm}{bx}{n}
\SetMathAlphabet{\mathit}{siptm}{OT1}{ptm}{m}{it}

\makeatletter
\def\@email#1#2{%
 \endgroup
 \patchcmd{\titleblock@produce}
  {\frontmatter@RRAPformat}
  {\frontmatter@RRAPformat{\produce@RRAP{*#1\href{mailto:#2}{#2}}}\frontmatter@RRAPformat}
  {}{}
}%
\makeatother

\makeatletter
\let\ps@titlepage\ps@plain
\makeatother

\newcommand{\bestmae}[1]{\textbf{#1}}
\newcommand*{\fref}[1]{Fig.~\ref{#1}}
\newcommand*{\tref}[1]{Table~\ref{#1}}
\newcommand*{\eref}[1]{Eq.~\eqref{#1}}
\newcommand*{\sref}[1]{Section~\ref{#1}}
\newcommand{\sifull}{Supplementary Information\xspace}
\newcommand{\sishort}{SI\xspace}

\newcommand{\epsxc}{\varepsilon_{\mathrm{xc}}}
\newcommand{\epsx}{\varepsilon_x}
\newcommand{\epsc}{\varepsilon_c}
\newcommand{\alphareg}{\alpha^{\mathrm{r2}}}
\newcommand{\code}[1]{\texttt{\detokenize{#1}}}
\begin{document}
\makeatletter
\def\@bibunitname{main_ref}
\makeatother
\begin{bibunit}
\global\let\mastertitle\title
\global\let\masterauthor\author
\global\let\masteraffiliation\affiliation
\global\let\masteremail\email
\global\let\masterdate\date
\global\let\mastermaketitle\maketitle
\global\let\masteraddcontentsline\addcontentsline
\renewcommand{\addcontentsline}[3]{}

\title[Agentic AI for Density-Functional Development]{%
Agentic AI for Density-Functional Development: Revisiting r2SCAN}

\author{Santosh Adhikari}
\affiliation{PsiQuantum, 700 Hansen Way, Palo Alto, California 94304, USA}
\email{sadhikari@psiquantum.com}
\author{Kelsey A. Parker}
\affiliation{PsiQuantum, 700 Hansen Way, Palo Alto, California 94304, USA}
\author{Etinosa Osaro}
\affiliation{PsiQuantum, 700 Hansen Way, Palo Alto, California 94304, USA}
\author{Swagata Roy}
\affiliation{PsiQuantum, 700 Hansen Way, Palo Alto, California 94304, USA}
\author{Dario Rocca}
\affiliation{PsiQuantum, 700 Hansen Way, Palo Alto, California 94304, USA}
\email{drocca@psiquantum.com}
\date{\today}

\begin{abstract}
In this work, we demonstrate physics-constrained agentic development of a meta-generalized gradient approximation (meta-GGA) functional using a large language model (LLM) to assist the search and optimization of a band-gap-oriented revision of r2SCAN.
Throughout optimization, no real bonded systems were fitted, preserving r2SCAN's nonempirical philosophy.
We externally validated three finalists from the agentic search.
Across the finalist set, band gaps and several molecular energetic and structural subsets improve relative to r2SCAN; the first-ranked finalist, r2SCAN+, reduces the band-gap MAE on a benchmark comprising 24 solids from 1.26 to 0.96 eV and the aggregate MAE across a benchmark comprising 329 reactions from 4.79 to 4.42 kcal/mol. 
To achieve this, we first curated 78 exchange and 90 correlation candidate correction terms from r2SCAN's dimensionless ingredients, spanning polynomial terms through third degree, exponentials, exponentially damped products, and ratios.
Even within this controlled search space, allowing each candidate to combine one to three correction terms from the exchange catalog, the correlation catalog, or both yields about \(8 \times 10^5\) distinct forms, making exhaustive high-throughput screening impractical.
We therefore used an LLM agent to prioritize promising combinations.
We defined its search criteria using r2SCAN's exact constraints, physical norms, and the targeted iso-orbital derivative response.
The agent then combined these criteria with its pretrained knowledge and accumulated search feedback to propose and refine sparse forms, prioritizing terms tied to the iso-orbital response; a second LLM critic screened proposals before deterministic verification.
Compared with uniform random search, the workflow learned from prior evaluations, incurred far fewer downstream rejections (0.6\% versus 24.6\%), and located stronger high-response candidates: 51 agentic candidates exceeded the best random-search response of 1.263, with the overall best reaching 1.331.
Overall, these results show that agentic search can support density-functional development when flexible hypothesis generation is coupled to automated physical verification.
\end{abstract}

\maketitle

\thispagestyle{plain}
\pagestyle{plain}

\section{Introduction}
\label{sec:introduction}

Density-functional theory (DFT)~\citep{hohenberg1964,kohn1965} is widely used in chemistry and materials science~\citep{jones2015density,maurer2019advances}.
Over the years, DFT has helped build an extensive infrastructure for scientific research and innovation: open-source software such as PySCF~\citep{sun2018pyscf,sun2020pyscf} and Quantum ESPRESSO~\citep{giannozzi2009,giannozzi2017,giannozzi2020}, associated libraries such as \textsc{libxc}~\citep{lehtola2018}, reusable benchmark and reference repositories~\citep{liang2025gscdb137,liang2025gscdbgithub,grossmann_mbpt,pederson2023dftexconditions} and a vast literature rich in physical insight and methodological ideas~\citep{teale2022dft,burke2012perspective,cohen2012challenges,becke2014fifty,perdew2013climbing,kaplan2023predictive,pederson2023exactconditions}.
This mature ecosystem provides not only the tools for applying existing density-functional approximations, but also much of the knowledge and computational machinery needed to develop new ones.
Exploiting these resources for functional development, however, requires researchers to connect theoretical ideas with analytic constructions, implementations, numerical tests, and validation procedures distributed across the literature and software.
Rapid advances in large language models (LLMs)~\citep{rein2024gpqa,epoch2025selfreportedgpqa,vals2026gpqa} suggest that LLM agents could assist this process under human scientific guidance by navigating these resources and coordinating reasoning with computation.

Recent studies indeed have shown that LLM agents can solve chemistry problems~\citep{bran2024chemcrow,boiko2023coscientist,gottweis2026coscientist,burger2020mobile,macleod2020selfdriving,szymanski2023alab}.
For example, Bran et al.\ used ChemCrow to autonomously plan and execute the syntheses of an insect repellent and three organocatalysts~\citep{bran2024chemcrow}.
Boiko et al.\ demonstrated autonomous experimental design and execution, including optimization of palladium-catalyzed cross-coupling reactions~\citep{boiko2023coscientist}.
Most recently, Google DeepMind and collaborators demonstrated a multi-agent system that iteratively generates, critiques, and refines scientific hypotheses, with experimental validation in biomedical applications~\citep{gottweis2026coscientist}.
Similarly, MLIPilot~\citep{osaro2026mlipilot}, following the closed-loop agentic workflow introduced by Karpathy~\citep{karpathy2026autoresearch}, demonstrates its application to atomistic machine learning.
Despite decades of accumulated literature and well-documented computational infrastructure, agentic applications remain scarce in the DFT community.

The ideal use of agentic artificial intelligence (AI) in DFT would be to assist the development of improved exchange--correlation approximations.
Functional construction broadly follows three routes: optimization against benchmark properties (empirical)~\citep{zhao2008m06,peverati2014quest}, construction from exact constraints and physical norms without fitting to properties of bonded systems (nonempirical)~\citep{pbe1996,pbesol2008,sun2015scan,furness2020r2scan}, or a mixture of these strategies (semi-empirical)~\citep{becke1997systematic,grimme2006b97d,brown2021mcml}.
Regardless of the route and the advances made within it, functional development remains laborious and time-consuming because researchers must acquire and synthesize substantial domain knowledge distributed across decades of literature and computational infrastructure, then perform the associated scientific reasoning and computational operations.
LLM agents can now assist with much of this intellectual integration.
In this work, we demonstrate this division of labor through a prototype agentic revision of a functional at the meta-generalized-gradient approximation (meta-GGA) level~\citep{perdew2001jacob}: we define the essential scientific objectives and admissible physical boundaries, while an LLM agent reasons over candidate forms, uses accumulated feedback to prioritize promising hypotheses, diagnoses failures, and coordinates coefficient optimization.

Meta-GGAs are especially attractive for such development because the kinetic-energy density provides flexibility to distinguish density environments that lower semilocal rungs cannot.
Nonempirical meta-GGAs such as SCAN~\citep{sun2015scan} and r2SCAN~\citep{furness2020r2scan} consequently provide broadly useful molecular and solid-state energetics and structures at semilocal cost~\citep{kingsbury2022r2scan}.
Their generalized-Kohn--Sham band gaps, however, remain substantially underestimated for many semiconductors and insulators~\citep{borlido2019bandgaps,borlido2020bandgaps}.
Reliable band structures are particularly relevant to optoelectronic engineering~\citep{schubert2005solidstate,polman2016photovoltaic} and catalytic applications~\citep{pinaud2011birnessite,adhikari2023hse06bands}.

Many-body (GW) calculations~\citep{hedin1965gw} and screened hybrids such as HSE06~\citep{heyd2003hse,krukau2006hse06} often improve gaps, but at substantially greater cost than semilocal DFT.
The semilocal mBJ potential~\citep{tran2009mbj,karolewski2013becke} can also yield accurate solid-state gaps, but as a model potential without a corresponding exchange--correlation energy functional it does not consistently provide total energies, forces, or structural optimization.
TASK~\citep{aschebrock2019ultranonlocality,lebeda2023rightgaps} instead showed that strengthening the response of meta-GGA exchange to the iso-orbital indicator can enlarge bulk band gaps at semilocal cost; mTASK~\citep{neupane2021mtask} extended this strategy to low-dimensional materials while explicitly relaxing the tight one-/two-electron exchange bound.
These studies established a useful band-gap control knob, although TASK and mTASK use PW92-LDA correlation~\citep{perdew1992pw92} and were not developed as broad replacements for the complete SCAN exchange--correlation construction~\citep{aschebrock2019ultranonlocality,lebeda2022binding}.

The recent LAK construction~\citep{lebeda2024lak,lebeda2025lak} demonstrates how much further this flexible space can be taken.
By rebalancing density-gradient and kinetic-energy-density contributions, LAK retains all 17 exact constraints available to a meta-GGA, achieves HSE06-like band-gap accuracy, and modestly improves a selected atomization-energy set over SCAN.
Its LC20 lattice-constant error is nevertheless comparable to PBE~\citep{pbe1996} rather than SCAN.
Thus, meta-GGA ingredients can balance band-gap response and molecular energetics within a fully semilocal form, but navigating the associated property trade-offs remains a demanding design problem.

In this work, we use physics-constrained agentic functional development to improve upon r2SCAN.
Although r2SCAN misses one of the 17 exact constraints satisfied by SCAN, it preserves SCAN's broad accuracy while providing greater numerical stability~\citep{furness2020r2scan,kingsbury2022r2scan}.
Our agentic workflow uses a human--agent interface: we define the scientific objective, admissible ingredients, physical tests, and acceptance rules, while the LLM agent uses pretrained knowledge, accumulated feedback, and an independent critique to select sparse exchange and correlation corrections, determine where they should act, and direct coefficient optimization.
The agent thus handles part of the reasoning and decision-making, including hypothesis selection, search steering, and numerical optimization, to assist the research process.
To assess the advantages and limitations of this agentic search, we also evaluated a uniform random-search control within the same constrained workflow.

Here, we follow a nonempirical route where no real bonded molecule or periodic solid enters the search objective, and candidates are judged only by exact-constraint verification, physical norms, and the iso-orbital response targeted for band-gap enhancement.
Remarkably, even within a candidate space deliberately restricted for simplicity and computational efficiency, the resulting agent-assisted functional, r2SCAN+, improves several external molecular subsets, suggesting that preserving r2SCAN's exact-constraint structure supports transferability beyond the band-gap target.
Its principal trade-off is PBE-like lattice-constant accuracy.
While this work was in preparation, Duston \textit{et al.}~\citep{duston2026agentic} reported a complementary agentic search around the range-separated hybrid functional $\omega$B97M-V~\citep{mardirossian2016wb97mv}, in which candidate functionals were evaluated non-self-consistently on fixed $\omega$B97M-V densities.
Their workflow used molecular benchmark data directly during optimization, whereas ours keeps benchmark molecules and solids outside the search loop and retains a fully nonempirical, semilocal meta-GGA framework.
Consistent with recent meta-GGA optimization studies~\citep{lebeda2024lak,lebeda2025lak}, our results show that agentic AI can support functional optimization: researchers retain control of the governing physics, while agents share the scientific reasoning and carry much of the computational optimization.

\section{Physics-Constrained Agentic Workflow}
\label{sec:workflow}

The workflow integrates the policy we define with an LLM proposer, an LLM critic, and deterministic verification.
We set the objective, ingredients, physical tests, and acceptance rules; the LLM agent, implemented here using the OpenAI GPT-5 reasoning model with medium reasoning effort~\citep{openai2025gpt5}, selects forms and directs optimization, while the verifier decides feasibility.
Below, we give an overview of the three central development steps: candidate construction and physical verification, agentic search and optimization, and finalist selection.
For completeness, using the same workflow, we also evaluate roughly the same number of candidates through a uniform random selection process to highlight the strengths and limitations of agentic search.

\subsection{Candidate construction and physical verification}
\label{sec:workflow:candidates}

Candidate functionals begin from r2SCAN and may modify one or both channels,
\begin{equation}
 \varepsilon_x=\varepsilon_x^{\mathrm{r2SCAN}}+\Delta\varepsilon_x,
 \qquad
 \varepsilon_c=\varepsilon_c^{\mathrm{r2SCAN}}+\Delta\varepsilon_c.
 \label{eq:workflow:decomposition}
\end{equation}
In \eref{eq:workflow:decomposition}, \(\Delta\varepsilon_\nu\) is the correction to channel \(\nu=x,c\).
The corresponding total energies are \(E_x=\int n(\mathbf r)\varepsilon_x(\mathbf r)\,d\mathbf r\) and \(E_c=\int n(\mathbf r)\varepsilon_c(\mathbf r)\,d\mathbf r\).

The candidate construction and derivative-response analysis use the r2SCAN-regularized iso-orbital indicator \(\alpha^{\mathrm{r2}}=(\tau-\tau_W)/(\tau_{\mathrm{unif}}+10^{-3}\tau_W)\), where \(\tau\), \(\tau_W\), and \(\tau_{\mathrm{unif}}\) have their standard meta-GGA meanings (see \sifull\ (\sishort), Sec. S1.1, for terminology definitions).
The curated catalogs of candidates contain 78 exchange and 90 correlation terms built from bounded transforms of \((r_s,s,p,\alpha^{\mathrm{r2}},\zeta)\) using polynomial, exponential, exponentially damped product, and ratio forms.
Candidate \(K\) combines one to three terms across either or both channels, \(P_\nu^K=\sum_jc_{\nu j}^K\phi_{\nu j}\) (the admissible forms of \(\phi_{\nu j}\) are catalogued in \sishort, Sec.~S1.2 and Tables~S1--S2), with \(1\le m_{x,K}+m_{c,K}\le3\), where \(m_{\nu,K}\) counts the active terms in channel \(\nu\).
The selected algorithm determines the active coefficients \(c_{\nu j}^K\).
This deliberate curation and sparsity reduce cost and favor interpretable expressions.

To confine a correction while preserving key r2SCAN limits, every active term in a candidate is multiplied by the same gating function
\begin{equation}
 g_\alpha(\alpha^{\mathrm{r2}};\eta,\sigma)=
 \frac{\alpha^{\mathrm{r2}}}{\alpha^{\mathrm{r2}}+\eta}
 \frac{\alpha^{\mathrm{r2}}-1}
 {1+[(\alpha^{\mathrm{r2}}-1)/\sigma]^2}.
 \label{eq:workflow:gate}
\end{equation}
Using this gate, the exchange and correlation expressions in \eref{eq:workflow:decomposition} for candidate \(K\) are
\begin{equation}
 \begin{aligned}
 \varepsilon_x^K&=\varepsilon_x^{\mathrm{r2SCAN}}
 +\varepsilon_x^{\mathrm{unif}}
 g_\alpha(\alpha^{\mathrm{r2}};\eta_K,\sigma_K)P_x^K,\\
 \varepsilon_c^K&=\varepsilon_c^{\mathrm{r2SCAN}}
 +\bar r_s\left|\varepsilon_c^{\mathrm{r2SCAN}}\right|
 g_\alpha(\alpha^{\mathrm{r2}};\eta_K,\sigma_K)P_c^K.
 \end{aligned}
 \label{eq:workflow:channel-construction}
\end{equation}
Here, \(\bar r_s=r_s/(1+r_s)\).
The corresponding exchange--correlation energy per electron and enhancement factor are
\begin{equation}
 \varepsilon_{\mathrm{xc}}^K=
 \varepsilon_x^K+\varepsilon_c^K,
 \qquad
 F_{\mathrm{xc}}^K=
 \frac{\varepsilon_{\mathrm{xc}}^K}{\varepsilon_x^{\mathrm{unif}}},
 \qquad
 \varepsilon_x^{\mathrm{unif}}=-\frac{3}{4\pi}(3\pi^2n)^{1/3}.
 \label{eq:workflow:enhancement}
\end{equation}
The workflow constructs this complete candidate before verification and applies exchange-, correlation-, or combined exchange--correlation probes as required by each physical condition.
In \eref{eq:workflow:gate}, a \(\max(0,\alpha^{\mathrm{r2}})\) filter is applied to \(\alpha^{\mathrm{r2}}\), so negative values caused by numerical noise are treated as zero.
The gate vanishes at \(\alpha^{\mathrm{r2}}=0\) and 1, forcing the correction \(\Delta\varepsilon_\nu\) to zero in the one-orbital and slowly varying limits, respectively, and thereby recovering r2SCAN in both limits.
Because the gate parameters can themselves influence the response, we used the same fixed pair, \((\eta,\sigma)=(0.6,1.8)\), in the agentic and uniform-random searches, ensuring that differences between them arise from candidate selection rather than gate tuning.
Although these parameters could in principle be optimized, we selected them a priori and held them fixed throughout all reported experiments.

In summary, these choices define the candidate space presented to the LLM agent: the admissible grammar and sparsity, together with the gate architecture and parameter bounds.
The physical tests and acceptance criteria used to evaluate candidates are described next.

\subsection{Agentic search}
\label{sec:workflow:search}

Within the candidate space defined in \sref{sec:workflow:candidates}, the agentic search seeks corrections that increase the targeted iso-orbital response while preserving the physical behavior of r2SCAN (see step 2 below).
Two LLM agents operate in the loop: a proposer and a critic.
For each iteration, the workflow proceeds through the following steps.

\begin{enumerate}

\item \textit{Proposal and critique.}
The proposer selects a candidate form from the available space, coefficient bounds and starting values for each selected term, and one of four listed algorithms to search those bounds for optimal coefficients; the shared gate pair remains fixed at \((\eta,\sigma)=(0.6,1.8)\).
Specifically, available optimizers are a one-dimensional scan, Nelder--Mead~\citep{nelder1965simplex}, covariance-matrix adaptation~\citep{hansen2001cma}, and Bayesian optimization~\citep{snoek2012bayesian}.
Together, these choices fully specify the starting exchange, correlation, and combined exchange--correlation expressions in \eref{eq:workflow:channel-construction} and \eref{eq:workflow:enhancement}, allowing the workflow to perform the deterministic constraint precheck (see \sishort, Sec~S1.3, and Table~S3 for the list of constraints checked).
The critic then reviews the complete proposal together with these initial feasibility results.
The critic either rejects the proposal or allows it to proceed, possibly warning that the starting coefficient vector is infeasible.
Such a warning does not force rejection because the selected algorithm may locate a feasible vector elsewhere within the proposed bounds.
Across the production runs, the critic rejected 330 of 1,738 proposals.
The remaining 1,408 proposals advanced to coefficient optimization under the fixed gate used for both search modes.

\item \textit{Search objective.}
Coefficient optimization is guided by three quantities evaluated relative to r2SCAN.
First, a physics-motivated surrogate for band-gap opening: the response of \(F_{\mathrm{xc}}^K\) in \eref{eq:workflow:enhancement} to the iso-orbital indicator.
For functional \(K\), we define
\begin{equation}
D_{\alpha^{\mathrm{r2}}}^K=
\left\langle\left|
\frac{\partial F_{\mathrm{xc}}^K}{\partial\alpha^{\mathrm{r2}}}
\right|^2\right\rangle_{G_{\alpha^{\mathrm{r2}}}}^{1/2},
\qquad
R_{\alpha^{\mathrm{r2}}}^K=
\frac{D_{\alpha^{\mathrm{r2}}}^K}
{D_{\alpha^{\mathrm{r2}}}^{\mathrm{r2SCAN}}}.
\label{eq:workflow:response}
\end{equation}
The average samples \((r_s,s,\alpha^{\mathrm{r2}})\) at \(\zeta=0\), making \(D^K\) the root-mean-square response and \(R^K\) its ratio to r2SCAN.
This choice follows TASK~\citep{aschebrock2019ultranonlocality} and mTASK~\citep{neupane2021mtask}, where increasing the exchange response to the kinetic-energy-density-dependent indicator promotes band-gap opening; those functionals originally used the \(\alpha\)-independent PW92-LDA correlation~\citep{perdew1992pw92}.
In this work, since both r2SCAN channels depend on the indicator and either or both may be corrected here, we optimize the response of \(F_{\mathrm{xc}}\).
Second, the workflow evaluates four scaling relations retained by r2SCAN: exchange uniform and nonuniform scaling and correlation high- and low-density scaling.
We define \(\mathcal L_{\gamma}^K\) as the aggregate normalized deviation of candidate \(K\) from their exact analytical targets.
Because finite-grid asymptotic probes can retain small numerical residuals even for r2SCAN, we use its identically evaluated value as the baseline for interpreting each candidate's aggregate scaling deviation.
Third, we define \(\mathcal L_{\Delta F_{\mathrm{xc}}}^K\) as the root-mean-square departure of the candidate enhancement factor from r2SCAN over synthetic points representing slowly varying densities.
All three quantities are analytic and inexpensive to evaluate; the nonempirical reference norms introduced in step 4 are not computed during coefficient optimization (see \sishort, Sec.~S1.5, for details on the computation of \(R_{\alpha^{\mathrm{r2}}}^K\), \(\mathcal{L}_{\gamma}^K\), and \(\mathcal{L}_{\Delta F_{\mathrm{xc}}}^K\)).
These quantities are combined in the objective
\begin{equation}
\begin{aligned}
J={}&-R_{\alpha^{\mathrm{r2}}}^K
 +0.1\mathcal L_{\Delta F_{\mathrm{xc}}}^K\\
 &+50\max\!\left[
 0,\mathcal L_{\gamma}^K-\mathcal L_{\gamma}^{\mathrm{r2SCAN}}
 \right],
\end{aligned}
\label{eq:workflow:objective}
\end{equation}
evaluated for coefficient vectors that pass the deterministic screening, while vectors that fail it are instead assigned
\begin{equation}
J_{\mathrm{infeasible}}=10^{3}\left(1+n_{\mathrm{viol}}\right),
\label{eq:workflow:penalty}
\end{equation}
where \(n_{\mathrm{viol}}\) is the number of violated constraints.
We chose the weights 0.1 and 50 manually; they prioritize enhancement of the iso-orbital response while discouraging substantial departures from r2SCAN in the two auxiliary measures.
The third term of \eref{eq:workflow:objective} is a hinge that vanishes whenever the candidate's aggregate scaling deviation does not exceed that of r2SCAN.
Infeasible vectors are ranked by their number of violations rather than discarded, which steers the search toward feasible regions; because r2SCAN itself corresponds to \(J=-1\), the two branches are separated by roughly three orders of magnitude.
\item \textit{Coefficient optimization and constraint screening.}
For each critic-approved proposal, the LLM-selected algorithm minimizes \eref{eq:workflow:objective} over the coefficients \(c_{\nu j}^K\) of the active terms within the proposed bounds; the gate pair \((\eta,\sigma)\) is held fixed, so only these linear coefficients are varied.
The algorithms differ in how they generate trial coefficient vectors: the one-dimensional method evaluates a uniform grid followed by a bounded refinement, the simplex method generates trial points by geometric updates of a simplex, and the evolutionary and Bayesian methods draw them stochastically.
Every trial vector, however generated, undergoes the same deterministic screening and receives a value from \eref{eq:workflow:objective} or \eref{eq:workflow:penalty} accordingly; the accumulated values then guide the next search step.
Optimization continues until the selected algorithm satisfies its convergence criterion or exhausts its evaluation budget.
The policy specified a nominal 60 evaluations, while method-specific stopping and accounting produced 7--116 objective calls per optimized production proposal, with a mean of approximately 69.
At termination, the lowest-\(J\) vector encountered is retained; it is necessarily feasible whenever any feasible vector was encountered, given the separation between \eref{eq:workflow:objective} and \eref{eq:workflow:penalty}.
If no feasible vector is found, the proposal is excluded from the accepted candidate pool.
No feasible coefficient vector was found for 9 of the 1,408 optimized proposals.
Consequently, 1,399 proposals advanced to full physical verification.

\item \textit{Full physical verification.}
Because evaluating the full nonempirical norms is substantially more expensive, we promote only the lowest-\(J\) feasible coefficient vector for each proposed form to this stage.
This cost-saving choice treats the inner objective as a screening surrogate; it does not guarantee that another feasible coefficient vector for the same form would not achieve a lower full norm loss.
The deterministic verifier of this step first repeats the constraint checks and then computes the full norm loss.
The loss contains six categories: scaling, comprising the same four scaling measures used above; rare-gas exchange and rare-gas correlation~\citep{burke2014atomiccorrelation}; large-\(Z\) coefficients; compressed Ar$_2$~\citep{patkowski2005argon}; and fractional-charge linearity (see \sishort, Sec.~S1.4 for more details).
We did not include jellium surface formation energies as a separate candidate-level norm in this study, although they were used as appropriate norms in constructing r2SCAN~\citep{furness2020r2scan}.
For reference quantity \(i\) in category \(g\),
\begin{equation}
r_i=\frac{P_i-R_i}{s_i},\quad
M_g=\frac{\sum_{i\in g}w_i r_i^2}{\sum_{i\in g}w_i},\quad
\mathcal L_{\mathrm{norm}}=
\left(\frac{\sum_gW_gM_g}{\sum_gW_g}\right)^{1/2}.
\label{eq:workflow:norm-loss}
\end{equation}
Here \(P_i\), \(R_i\), and \(s_i\) are the prediction, reference, and fixed scale; \(r_i\) is the residual; \(w_i\) weights quantity \(i\); \(M_g\) is the category mean square; and \(W_g\) weights category \(g\) (see \sishort, Table S4, for complete residual scales and category weights used in the loss terms).
A candidate is accepted only if \(\mathcal L_{\mathrm{norm}}^K\le \mathcal L_{\mathrm{norm}}^{\mathrm{r2SCAN}}\).
For every accepted candidate, the workflow records its optimized coefficients, response, objective value, and norm loss.
To reduce this cost further, each promoted candidate's Ar$_2$ interaction energies are evaluated non-self-consistently on r2SCAN densities.

\item \textit{Feedback and repetition.}
The workflow records every accepted or rejected outcome, including critic decisions, optimized coefficients, objective values, feasibility failures, derivative responses, and norm losses.
For computational efficiency, the proposer generates forms in batches of up to ten.
The first batch shares the same initial information; before each subsequent batch, the workflow provides a compact summary of all preceding results, including the best candidates by norm loss and derivative response and the most recent rejections, so that the proposer can use this accumulated feedback to guide its next forms and search settings.
The sequence of proposal, critique, coefficient search, and verification repeats until convergence or the allocated compute or LLM-cost budget is reached.

\end{enumerate}

Throughout this workflow, no molecular or solid-state benchmark dataset enters candidate generation, coefficient optimization, or selection; these datasets are reserved for external validation of the finalists.
We fix the architecture, catalogs, sparsity, gates, parameter ranges, objective weights, grids, tolerances, and acceptance thresholds.
Within these boundaries, the proposer chooses forms and search settings, the critic challenges them, the selected algorithm determines the coefficients, and the deterministic verifier alone determines feasibility and final survival.

\subsection{Agentic finalists}
\label{sec:workflow:finalists}

Of the 1,399 proposals that reached full physical verification across the production runs, 1,150 met the full r2SCAN norm-loss budget.
We compared these accepted, agent-proposed candidates by their norm loss and derivative-response opening and retained the three candidates with the largest band-gap responses (\eref{eq:workflow:finalists}); all three were correlation-only by outcome.
Throughout the manuscript, we label these candidates r2SCAN+, Finalist 2, and Finalist 3 in decreasing order of derivative response.

Their correction and analytic expansions are
\begin{equation}
 \begin{aligned}
 \Delta\varepsilon_c^K&=
 \bar r_s|\varepsilon_c^{\mathrm{r2SCAN}}|
 g_\alpha(\alpha^{\mathrm{r2}};\eta_K,\sigma_K)P_c^K,\\
 P_c^{\mathrm{r2SCAN+}}&=0.236802e^{-\bar\alpha}
 +0.390769e^{-\bar\alpha^2}
 +0.498078e^{-\bar\alpha^3},\\
 P_c^{\mathrm{Finalist~2}}&=0.485686e^{-\bar\alpha^2}
 +0.495912e^{-\bar\alpha^3}
 +0.193335\bar r_s e^{-\bar r_s},\\
 P_c^{\mathrm{Finalist~3}}&=0.499610e^{-\bar\alpha}
 +0.499965e^{-\bar\alpha^3}
 +0.199782\frac{\bar r_s}{1+\bar s}.
 \end{aligned}
 \label{eq:workflow:finalists}
\end{equation}
In \eref{eq:workflow:finalists}, \(\bar s=s/(1+s)\), \(\bar p=s^2/(1+s^2)\), \(\bar\alpha=\alpha^{\mathrm{r2}}/(1+\alpha^{\mathrm{r2}})\), and \((\eta_K,\sigma_K)=(0.6,1.8)\).
The search did not identify an exchange-containing candidate with comparably large response; for illustration, an exchange-containing candidate reached a response of only \(\sim 1.22\).

\begin{figure}[h]
  \centering
  \includegraphics[width=\columnwidth]{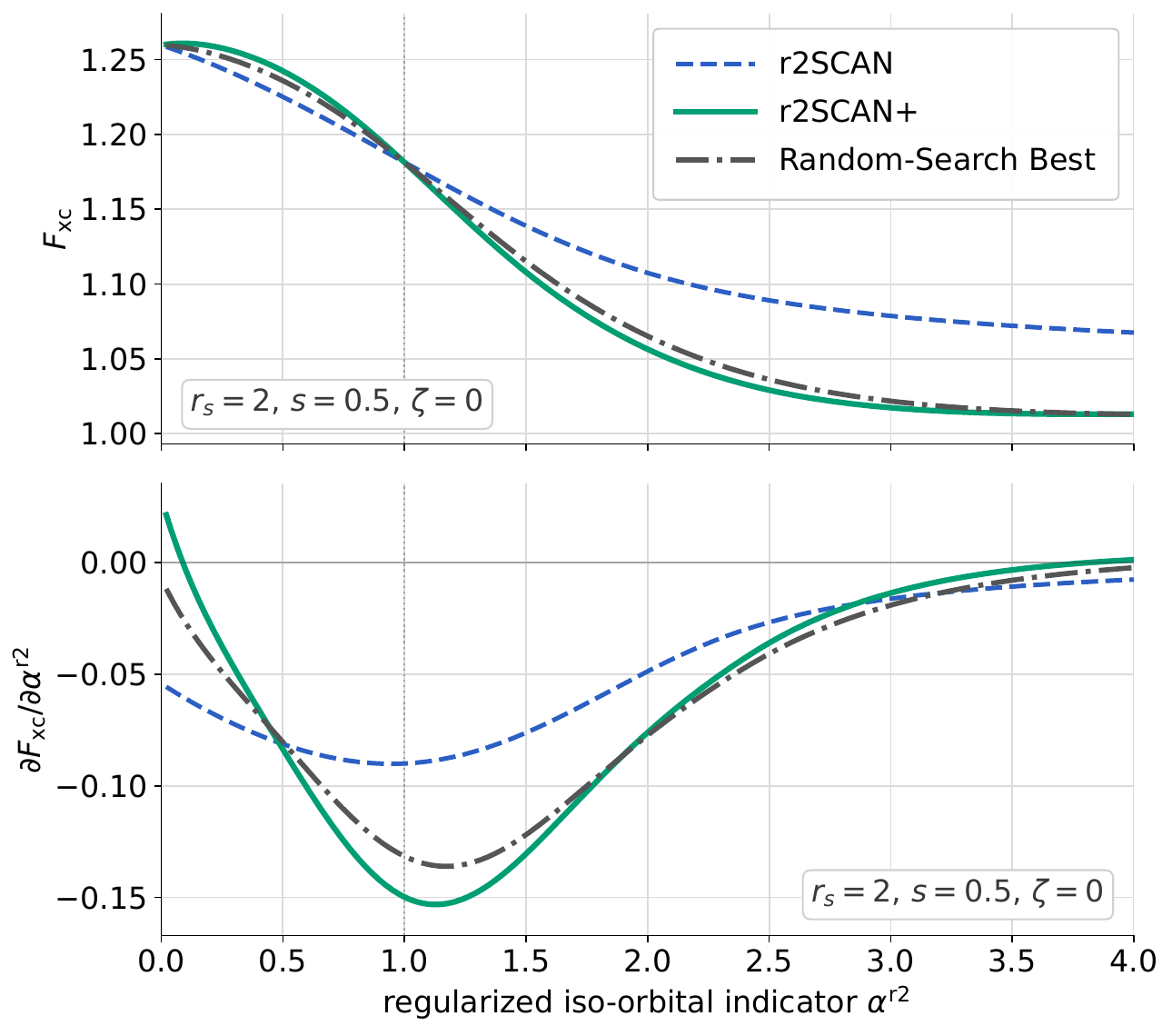}
  \caption{\(F_{\mathrm{xc}}\) and its signed \(\alpha^{\mathrm{r2}}\) derivative for r2SCAN, r2SCAN+, and Random-Search Best at the representative slice \(r_s=2.0\), \(s=0.5\), and \(\zeta=0\).
The full diagnostic-grid RMS derivative ratios are reported in \tref{tab:workflow:finalists}.}
  \label{fig:workflow:response}
\end{figure}

We summarize the finalists' full-grid norm losses and derivative-response gains in \tref{tab:workflow:finalists}.
\fref{fig:workflow:response} compares the smooth variation of \(F_{\mathrm{xc}}\) and its iso-orbital response for r2SCAN, r2SCAN+, and uniform Random-Search Best.
Additional response plots for representative \((\bar r_s,s,\zeta)\) slices are provided in Fig.~S1 of the \sishort.
The complete top-ten ranked lists and optimized expressions are given in the \sishort, Sec.~S2.1 and Tables~S5 and S6; the full search-funnel comparison is summarized in Table~S7.
\begin{table}[h]
\centering
\scriptsize
\setlength{\tabcolsep}{2.5pt}
\caption{Norm-loss gains relative to r2SCAN and derivative-response ratios for the three agentic finalists and three uniform random-search candidates; the r2SCAN reference is 0\% and 1.0, respectively.
Random-Search Best, Random-Search II, and Random-Search III are the first, second, and third best candidates from the uniform random search, respectively; detailed statistics are given in the \sishort.}
\label{tab:workflow:finalists}
\begin{tabular}{@{}lcl|lcl@{}}
\toprule
\multicolumn{3}{c|}{Agentic Search} & \multicolumn{3}{c}{Random Search} \\
Candidate & \shortstack{Norm-loss\\gain (\%)} & \(R_{\alpha^{\mathrm{r2}}}\) & Candidate & \shortstack{Norm-loss\\gain (\%)} & \(R_{\alpha^{\mathrm{r2}}}\) \\
\midrule
r2SCAN+ & 0.44 & 1.331 & Random-Search Best & 0.01 & 1.263 \\
Finalist 2 & 0.27 & 1.329 & Random-Search II & 1.11 & 1.261 \\
Finalist 3 & 0.41 & 1.328 & Random-Search III & 0.87 & 1.241 \\
\bottomrule
\end{tabular}
\end{table}

Among the finalists, r2SCAN+ gives the largest response, 33.1\% above r2SCAN.
We use r2SCAN+ for the main-text comparisons with SCAN and r2SCAN.

\subsection{Comparing Agentic and Uniform Random Search}
\label{sec:workflow:manual-comparison}

To assess the benefit of agent-guided candidate selection, we compared the agentic workflow with a uniform random search using the same 168-term menu, fixed gate \((\eta,\sigma)=(0.6,1.8)\), coefficient objective, physical constraints, norm-loss verifier, and proposal budget.
Across the different initializations, the random search supplied 2,480 proposals, compared with 1,738 proposals supplied by the agentic search.
Each search represented less than \(0.3\%\) of the available approximately \(8 \times 10^5\)-form combinatorial space.

The agentic workflow reached the r2SCAN norm-loss budget for 1,150 of 1,738 proposals (66.2\%), compared with 1,270 of 2,480 random-search proposals (51.2\%).
The agentic search also led to stronger responses: 110 agentic proposals exceeded \(R_{\alpha^{\mathrm{r2}}}=1.20\), compared with 18 random-search proposals.
The random-search best response of 1.263 was surpassed by 51 agentic candidates, while the best agentic candidate, r2SCAN+ in \eref{eq:workflow:finalists}, achieved a response of 1.331.
The norm-loss gains and opening responses of the three random-search candidates are listed in \tref{tab:workflow:finalists}.

Examining the search candidates, the agentic arm included an explicit \(\alpha^{\mathrm{r2}}\)-dependent term in \(70.7\%\) of proposals, compared with \(32.6\%\) under uniform random sampling, consistent with the objective of increasing the response with respect to \(\alpha^{\mathrm{r2}}\).
The searches also differed in the distribution of correction types: exchange-only, correlation-only, and joint forms accounted for \(12.7\%\), \(43.6\%\), and \(43.7\%\) of the agentic proposals, respectively, compared with \(9.8\%\), \(15.9\%\), and \(74.3\%\) of the random-search proposals.
Together, these patterns are consistent with the agent using previous evaluations to guide subsequent proposals toward \(\alpha\)-dependent correlation corrections that enhance the band-gap response.
All three finalists, for example, contain exponential \(\bar{\alpha}\)-dependent terms: r2SCAN+ includes \(e^{-\bar{\alpha}}\), \(e^{-\bar{\alpha}^2}\), and \(e^{-\bar{\alpha}^3}\), while Finalists~2 and~3 each include two of these terms (\eref{eq:workflow:finalists}).
This pattern extends beyond the finalists: all ten highest-response agentic candidates contain at least one of these $\bar{\alpha}$-dependent exponential terms.
By contrast, none of the ten highest-response random-search candidates contains an explicit \(\alpha\)-dependent term, including the random-search candidate with the highest response.
Thus, the random-search response enhancement arose from the shared gate architecture rather than deliberate selection of \(\alpha\)-dependent correction terms; the random search did not use the gate's role as a criterion when selecting candidate forms.

The agentic workflow also benefits from a critic agent that is absent from the random-search control.
Across the 1,738 agentic proposals, the critic rejected 330 before coefficient optimization, saving the associated optimization cost; among the 1,408 proposals that passed the critic, only 9 (0.6\%) were subsequently found structurally infeasible, while 1,150 (81.7\%) remained within the r2SCAN norm-loss budget.

We therefore identify three practical advantages of the agentic workflow: it prioritizes terms connected to the dominant loss objective, uses successive evaluations and log history to refine later proposals, and applies critic screening before coefficient optimization; consequently, 51 agentic candidates surpassed the best random-search response of 1.263, with r2SCAN+ reaching 1.331.
At the same time, the shared gate architecture and the common r2SCAN-based workflow keep both searches competitive, with 66.2\% of agentic proposals and 51.2\% of random-search proposals remaining within the r2SCAN norm-loss budget.
Since each search explored well below 0.5\% of the available combinatorial space, these results yet do not establish that agentic search guarantees an optimal candidate.
Rather, we find that its ability to make informed decisions, learn from accumulated logs, and follow user-defined rules makes agentic search an appealing and promising route for systematic functional exploration.
Throughout the remainder of the manuscript, we focus on the performance of the agentic finalists.
We also externally benchmarked the best random-search candidate; those results are reported in the \sishort and briefly compared with the agentic candidates in the Results section.

\section{Computational Details}
\label{sec:computational-details}

For external molecular validation, we selected a suite of 329 unique reactions from the most recent curated GSCDB137 dataset of Head-Gordon et al.~\citep{liang2025gscdb137,liang2025gscdbgithub}.
The suite comprises 183 non-multireference and 17 multireference W4-17 atomization energies~\citep{karton2017w417}; 68 unique barrier heights from DBH22~\citep{zheng2009dbh24} and BH46~\citep{goerigk2010gmtkn30}, with BH6~\citep{lynch2003ae6bh6} reported separately as a six-reaction representative subset of DBH22; and 25 G21EA electron affinities and 36 G21IP ionization potentials~\citep{curtiss1991g21,goerigk2010gmtkn30}.
We performed the calculations with PySCF 2.9.0~\citep{sun2018pyscf,sun2020pyscf}, using RKS for closed-shell species and UKS for open-shell species.
We used def2-QZVPPD~\citep{weigend2005def2,rappoport2010} basis set for W4-17 and the barrier-height sets and aug-cc-pV5Z~\citep{dunning1989,kendall1992} for G21EA and G21IP, except that species containing Li, Be, Na, or Mg used def2-QZVPPD.
For benchmark calculations, we used the PySCF level-6 integration grid, an SCF energy threshold of \(10^{-7}\) Hartree, and at most 200 cycles.
We evaluated SCAN, r2SCAN, and the three finalists self-consistently under these common settings.
To study numerical-integration sensitivity we also performed calculations for the complete molecular suite at grid levels 2--5.
For comparison, we also computed PBE~\citep{pbe1996} and B3LYP~\citep{becke1993,stephens1994b3lyp} using the same dataset-specific basis sets and computational settings.

We additionally optimized the 20 bond-lengths in MGBL20~\citep{zhao2008sogga,peverati2012xc}, comprising nine bonds involving hydrogen in MGHBL9~\citep{zhao2008sogga} and 11 bonds between heavier atoms in MGNHBL11~\citep{peverati2012xc}, distributed across 16 molecules.
We used the def2-QZVPPD basis set, PySCF level-6 grid, an SCF threshold of \(10^{-10}\) Hartree, analytic nuclear gradients with grid response, and the geomeTRIC optimizer~\citep{wang2016geometric}.

For periodic system validation, we considered the zero-point-corrected experimental lattice constants of the LC20 dataset reported in Ref.~\onlinecite{csonka2009}.
This set contains six simple metals, five semiconductors, five ionic crystals, and four late transition metals.
Additionally, we considered a separate nonmetallic benchmark of 24 unique solids spanning elemental and compound semiconductors, ionic insulators, and rare-gas solids.
It combines the 14-solid TASK benchmark~\citep{aschebrock2019ultranonlocality} and SCBG15, the 15-solid benchmark of elemental and binary semiconductors used for LAK~\citep{lebeda2024lak}; the two sets share five solids.
We took the experimental reference gaps from the MBPT benchmark repository wherever available~\citep{grossmann_mbpt,borlido2019bandgaps}; for MgS and CdO we retained the TASK values of \(5.40\) and \(0.84\) eV, respectively~\citep{aschebrock2019ultranonlocality}, while ZnTe follows the LAK reference~\citep{lebeda2024lak}.
We likewise used the HSE06 gaps compiled in the MBPT repository wherever available.
We supplied the missing MgS value, \(4.48\) eV, from Lucero et al.~\citep{lucero2012hiss}, the CdO value, \(0.76\) eV, from Kim et al.~\citep{kim2020bandgapdatabase}, and the ZnTe value, \(2.19\) eV, from LAK~\citep{lebeda2024lak}.

We performed the calculations with Quantum~ESPRESSO 7.5~\citep{giannozzi2009,giannozzi2017,giannozzi2020} and scalar-relativistic SG15 ONCV PBE norm-conserving pseudopotentials~\citep{hamann2013,schlipf2015sg15}.
For every material and functional, we used wave-function and charge-density cutoffs of 160 and 640 Ry, respectively, and \(\Gamma\)-centered Monkhorst--Pack meshes targeting a reciprocal-space spacing of \(0.10~\text{\AA}^{-1}\).
We converged the self-consistent calculations to \(10^{-9}\) Ry with a mixing parameter of 0.4 and a maximum of 200 electronic steps.
For metallic systems, we used Marzari--Vanderbilt (``mv'') smearing with a smearing width of \(0.01\) Ry; the small-gap Ge calculations used the same smearing with a width of \(0.005\) Ry.

To determine each LC20 lattice constant, we sampled nine isotropically scaled cells spanning \(\pm4\%\) about the zero-point-corrected experimental reference lattice constant reported in Ref.~\onlinecite{csonka2009}.
We fitted a third-order Birch--Murnaghan equation of state~\citep{birch1947eos}.
For the band gaps, we used the experimental MBPT/ICSD structures wherever available.
For zincblende MgS and rocksalt CdO, we used the crystallographic phases and experimental lattice constants reported in TASK; for zincblende ZnTe, we used those reported in LAK.

We generated primitive cells and high-symmetry paths with SeeK-path~\citep{hinuma2017}, included 16 unoccupied bands, and omitted spin--orbit coupling.
We used identical structures, pseudopotentials, cutoffs, and sampling settings for all directly compared functionals.

\section{Results}
\label{sec:results}

\begin{figure*}[!t]
  \centering
  \includegraphics[width=0.92\linewidth]{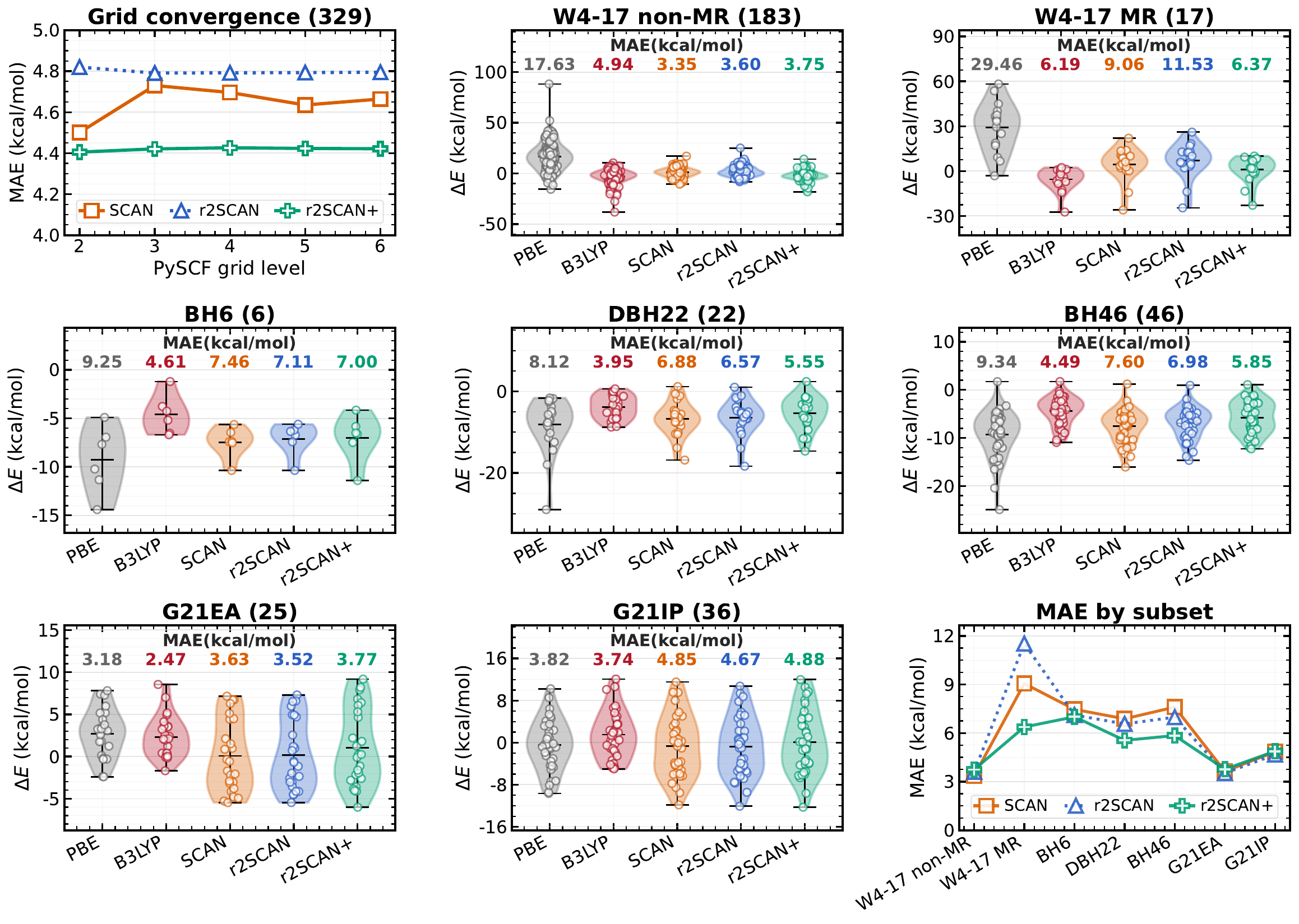}
  \caption{Molecular reaction-energy validation.
    The first panel shows the overall MAE of SCAN, r2SCAN, and r2SCAN+ across PySCF grid levels 2--6.
    The next seven panels show the level-6 signed-error distributions and subset MAEs for PBE, B3LYP, SCAN, r2SCAN, and r2SCAN+.
    The final panel compares the level-6 meta-GGA MAEs by subset.}
  \label{fig:results:molecular}
\end{figure*}

\begin{figure}[!t]
  \centering
  \includegraphics[width=\columnwidth]{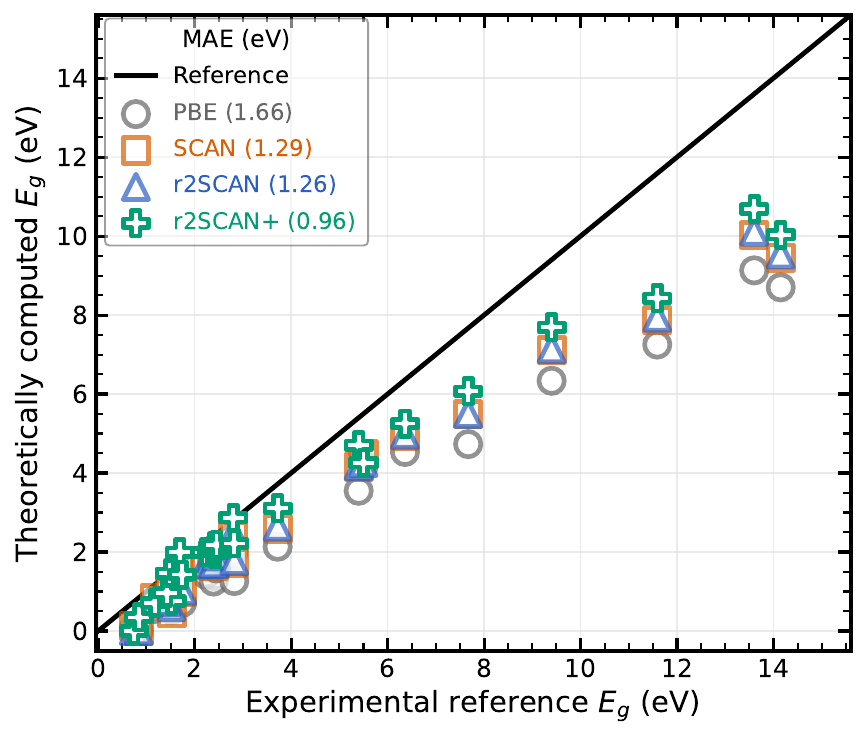}
  \caption{Fundamental gaps for the 24 unique solids in the union of TASK14~\citep{aschebrock2019ultranonlocality} and SCBG15~\citep{lebeda2024lak}.
    Legend values are MAEs in eV over this union for the methods computed here.
    On TASK14, the reference HSE06 MAE compiled from values available in the literature is \(1.02\) eV~\citep{grossmann_mbpt,borlido2019bandgaps,lucero2012hiss,kim2020bandgapdatabase}, while published TASK and mTASK predictions give \(0.39\) and \(0.47\) eV, respectively~\citep{aschebrock2019ultranonlocality,neupane2021mtask}.
    On SCBG15, the reference HSE06 MAE ranges from \(0.15\) to \(0.17\) eV~\citep{grossmann_mbpt,borlido2019bandgaps,lebeda2024lak}.}
  \label{fig:results:bandgaps}
\end{figure}

Because the finalists were developed nonempirically, without fitting to molecular or solid-state benchmark data, we next assess their transferability across the molecular and periodic validation datasets described in \sref{sec:computational-details}.
In this section, we focus on r2SCAN+, comparing it primarily with its parent functional, r2SCAN, and with SCAN, while placing the results in the context of the broader literature.
\sishort Sec.~S3.2 and Tables~S8--S11 provide all numerical values underlying the molecular-reaction, bond-length, band-gap, and lattice-constant analyses for the agentic finalists and Random-Search Best; the grid-convergence table in \sishort Sec.~S3.1 reports subset and overall MAEs across PySCF grid levels 2--6, and Table~S12 summarizes the benchmark comparison.
For a brief overview, \tref{tab:results:summary} summarizes performance across the molecular and periodic benchmark groups, while \fref{fig:results:molecular} and \fref{fig:results:bandgaps} show the detailed molecular and band-gap results, respectively.

First, we report that all molecular calculations associated with the 329 unique reaction energies completed without an SCF convergence failure for SCAN, r2SCAN, r2SCAN+, Finalist 2, and Finalist 3.
More importantly, like its parent r2SCAN, r2SCAN+ changes by about \(0.02\) kcal/mol in overall MAE across grid levels 2--6, whereas SCAN shows the largest drift, about \(0.23\) kcal/mol across levels 2--6 (top-left panel of \fref{fig:results:molecular}).
The numerical sensitivity of SCAN was a major motivation for the regularization introduced in r2SCAN~\citep{furness2020r2scan}.
Our results show that r2SCAN+ and the other agentic finalists retain the smooth grid behavior of r2SCAN.

Across all 329 molecular energetic data points, r2SCAN+ gives an MAE of \(4.42\) kcal/mol, close to the \(4.56\) kcal/mol MAE of B3LYP~\citep{becke1993,stephens1994b3lyp}; SCAN and r2SCAN give \(4.66\) and \(4.79\) kcal/mol, respectively (\tref{tab:results:summary}).
Relative to r2SCAN, r2SCAN+ reduces the errors in atomization energies and barrier heights across every evaluated subset.
The non-multireference and multireference W4-17 MAEs change from \(3.60\) and \(11.53\) to \(3.75\) and \(6.38\) kcal/mol, respectively, lowering the aggregate W4-17 MAE from \(4.28\) to \(3.97\) kcal/mol.
The BH6, DBH22, and BH46 MAEs decrease from \(7.11\), \(6.57\), and \(6.98\) to \(7.00\), \(5.55\), and \(5.85\) kcal/mol.
G21EA and G21IP remain close to r2SCAN, changing only from \(3.52\) to \(3.77\) and from \(4.67\) to \(4.88\) kcal/mol.
Thus, the molecular results show broad improvement without a noticeable degradation in any evaluated category; \fref{fig:results:molecular} provides the detailed subset distributions.
Because none of the 329 molecular benchmark data points entered the candidate-generation or selection procedure, these improvements provide encouraging evidence that preserving exact constraints and physical norms can support transferability beyond the targeted band-gap response.

Bond-length accuracy also shows transferability at the geometry level.
On MGBL20, r2SCAN+ gives an MAE of \(3.3\)~m\AA{}, compared with \(3.8\)~m\AA{} for r2SCAN.
The MGHBL9 MAE changes from \(2.0\) to \(2.2\)~m\AA{}, whereas the MGNHBL11 MAE improves from \(5.2\) to \(4.3\)~m\AA{}.
r2SCAN+ therefore remains comparatively balanced between hydrogen-involving and heavy-atom bonds, broadly following r2SCAN and avoiding the larger SCAN subset contrast of \(1.8\) versus \(6.4\)~m\AA{}.

\noindent\begin{minipage}{\linewidth}
\centering
\scriptsize
\setlength{\tabcolsep}{2.5pt}
\captionof{table}{MAEs for broad subsets of the external validation suite.
  Boldface marks the lowest MAE in each row.
  The TASK14 and SCBG15 gap sets share five solids; asterisks mark these overlapping sets and their 24-system union, in which each solid is counted once.
  At the aggregate category level, r2SCAN+ performs best among the methods evaluated here for reaction energies, bond lengths, and band gaps; lattice constants are the exception.}
\label{tab:results:summary}
\begin{tabular}{@{}lrrrr@{}}
\toprule
External validation set (\(N\)) & PBE & SCAN & r2SCAN & r2SCAN+ \\
\midrule
\multicolumn{5}{l}{\emph{Molecular thermochemistry and kinetics (kcal/mol)}} \\
Atomization energies (200) & 18.63 & \bestmae{3.84} & 4.28 & 3.97 \\
Barrier heights (68)        &  8.95 & 7.37 & 6.85 & \bestmae{5.75} \\
G21EA and G21IP (61)        & \bestmae{3.56} & 4.35 & 4.20 & 4.42 \\
Overall (329)               & 13.84 & 4.66 & 4.79 & \bestmae{4.42} \\
\midrule
\multicolumn{5}{l}{\emph{MGBL20 bond lengths (m\AA)}} \\
MGHBL9 (9)                  & 10.6 & \bestmae{1.6} & 2.0 & 2.2 \\
MGNHBL11 (11)               &  6.2 & 6.8 & 5.2 & \bestmae{4.3} \\
Overall (20)                &  8.2 & 4.4 & 3.8 & \bestmae{3.3} \\
\midrule
\multicolumn{5}{l}{\emph{LC20 lattice constants (m\AA)}} \\
All metals (10)             & 35.1 & \bestmae{22.3} & 26.4 & 85.8 \\
Insulators and semiconductors (10)
                            & 89.1 & \bestmae{25.9} & 31.2 & 63.6 \\
Overall (20)                & 62.1 & \bestmae{24.1} & 28.8 & 74.7 \\
\midrule
\multicolumn{5}{l}{\emph{Fundamental band gaps (eV)}} \\
TASK14 (14)$^*$             & 2.24 & 1.79 & 1.78 & \bestmae{1.43} \\
SCBG15 (15)$^*$             & 0.92 & 0.66 & 0.63 & \bestmae{0.39} \\
Overall (24)$^*$            & 1.68 & 1.29 & 1.26 & \bestmae{0.96} \\
\bottomrule
\end{tabular}
\end{minipage}

Next, we compare band gaps across 24 unique solids: the union of the 14-solid TASK benchmark~\citep{aschebrock2019ultranonlocality,neupane2021mtask} and the 15-solid SCBG15 benchmark used for LAK~\citep{lebeda2024lak}, with their five shared systems counted once.
Across this union, r2SCAN+ lowers the r2SCAN MAE from \(1.26\) to \(0.96\) eV; SCAN and PBE give \(1.29\) and \(1.66\) eV, respectively (\fref{fig:results:bandgaps}).
This improvement is consistent with our agentic design objective: increasing the response of \(F_{xc}\) to the regularized iso-orbital indicator, an established meta-GGA control knob for increasing generalized-Kohn--Sham ultranonlocality and promoting band-gap opening.

On TASK14, r2SCAN+ lowers the r2SCAN MAE from \(1.78\) to \(1.43\) eV.
Although this is a systematic improvement, the published TASK and mTASK MAEs of \(0.39\) and \(0.47\) eV, respectively, are substantially lower on this set~\citep{aschebrock2019ultranonlocality,neupane2021mtask}.
The central difference lies in the design objective: TASK and mTASK were developed specifically to improve band gaps, whereas we ask whether an LLM agent can identify useful corrections within a restricted neighborhood of r2SCAN while retaining its exact constraints, regularized architecture, and physical-norm performance.
The simultaneous improvement in band gaps and most external molecular subsets supports this broader balance, which was not tested for molecular transferability in the original TASK and mTASK studies.

Recent work on the more general optimization of meta-GGAs provides a complementary reference~\citep{lebeda2022binding,lebeda2024lak,lebeda2025lak}.
LAK~\citep{lebeda2024lak} demonstrates that stronger band-gap response can coexist with accurate atomization energies and bond lengths while retaining exact constraints.
On SCBG15, LAK reports an MAE of \(0.18\) eV, close to the HSE06 range of \(0.15\)--\(0.17\) eV compiled from the literature~\citep{grossmann_mbpt,borlido2019bandgaps,lebeda2024lak}.
r2SCAN+ improves the r2SCAN MAE from \(0.63\) to \(0.39\) eV but remains less accurate than LAK for this set.
For the MGBL20 bond-length set, however, r2SCAN+ and LAK give similar MAEs of \(3.3\) and \(3.4\)~m\AA{}, respectively, both slightly better than \(3.8\)~m\AA{} for r2SCAN and \(4.4\)~m\AA{} for SCAN.

Here, we focus on r2SCAN+ because it gives the largest derivative-response opening, the designated target of our search.
Its \(33.1\%\) response gain yields a 24-solid band-gap MAE of \(0.96\)~eV.
Finalists~2 and~3 perform similarly to r2SCAN+, consistent with their closely related correlation corrections and derivative-response gains of \(32.9\%\) and \(32.8\%\), respectively.
Their 24-solid band-gap MAEs are both \(0.96\)~eV, with molecular reaction MAEs of \(4.44\) and \(4.42\)~kcal/mol, respectively, compared with \(4.42\)~kcal/mol for r2SCAN+.
Random-Search Best has a slightly smaller derivative response and correspondingly a slightly higher band-gap MAE of \(1.02\)~eV, while giving a molecular MAE of \(4.37\)~kcal/mol.
These improvements in band gaps and molecular energetics come with a common structural trade-off: for the LC20 lattice-constant benchmark MAEs are \(0.075\), \(0.075\), \(0.077\), and \(0.069\)~\AA{} for r2SCAN+, Finalists~2 and~3, and Random-Search Best, respectively, compared with \(0.062\)~\AA{} for PBE (\tref{tab:results:summary}).

The nonempirical LAK functional~\citep{lebeda2024lak}, designed to improve band gaps and electronic binding simultaneously, exhibits a similar structural trade-off. 
By contrast, the recently introduced bn-r2SCAN functional~\citep{wang2026balanced} achieves a notably better balance among band gaps, lattice constants, and several molecular properties, although it yields somewhat less accurate atomization energies than r2SCAN. 
Consistent with our findings, its analysis shows that the correlation channel, in addition to exchange, can actively contribute to band-gap improvement. 
This balanced performance is obtained by fitting bn-r2SCAN to real bonded systems for both band gaps and lattice constants, whereas our work adopts a fully nonempirical fitting strategy. 

Our purpose here is not to identify an optimal functional, as any such claim would require broader molecular and materials validation.
Moreover, even our combined agentic and uniform random searches explore only a small fraction of the available space, leaving substantial opportunities for further development.
Together with recent meta-GGA optimization studies~\citep{lebeda2022binding,lebeda2024lak,lebeda2025lak,wang2026balanced}, our results suggest that this unexplored space offers considerable scope to rebalance electronic, structural, and molecular properties.
For example, future searches could explicitly balance the iso-orbital response targeted here for band-gap opening with the reduced-gradient behavior important for lattice constants, as established by previous studies~\citep{pbesol2008,csonka2009,lebeda2024lak,wang2026balanced}.
More broadly, our results show that agentic search provides a promising route to explore this space more systematically, by combining physical reasoning, accumulated search feedback, and numerical optimization.

\section{Conclusion}
\label{sec:conclusion}

In this work, we demonstrated a physics-constrained agentic workflow for developing density-functional approximations, using correction on r2SCAN as a prototype.
The resulting r2SCAN+ reduces the 24-solid band-gap MAE from \(1.26\) eV for r2SCAN to \(0.96\) eV and gives an overall 329-reaction MAE of \(4.42\) kcal/mol, with notable gains for several atomization and barrier-height subsets.
Because the external molecular benchmark sets did not enter candidate generation, optimization, or selection, these gains provide encouraging evidence that preserving exact constraints and physical norms can support transferability beyond the targeted band-gap response.

The molecular gains are broad, with electron affinities and ionization potentials remaining close to r2SCAN; the principal trade-off is the larger LC20 lattice-constant error, particularly for simple-metal and ionic solids.
r2SCAN+ also does not yet reach the band-gap accuracy of meta-GGAs optimized specifically for that property.
Nevertheless, its simultaneous improvements across band gaps, atomization energies, and barrier heights show that a useful balance can be found around the regularized r2SCAN architecture.
Wider molecular and materials benchmarks, including noncovalent, magnetic, and strongly correlated systems, will be needed to establish the generality of this balance and guide further refinement toward a more broadly improved meta-GGA.

We also compared the agentic workflow with uniform random search within the same constrained setting.
Unlike random exploration, the agent uses the search objective, physical rules, critic review, and previous outcomes to guide its proposals.
This is reflected in its preference for \(\alpha\)-dependent terms, iterative addition and replacement of candidate terms, a substantially lower rejection rate during constraint checking (\(0.6\%\) versus \(24.6\%\)), and stronger responses: 51 agentic candidates exceeded the best response obtained by uniform random search (1.26), while the best agentic candidate, r2SCAN+, reached 1.33.
At the same time, the common gate already enforces important physical behavior and introduces \(\alpha\) dependence in both searches, allowing random forms to remain competitive.
Random-Search Best, for example, contains no explicit \(\alpha\)-dependent correction term yet gives MAEs of \(4.37\)~kcal/mol for the 329-reaction molecular set, \(1.02\)~eV for the 24-solid band gaps, and \(0.069\)~\AA{} for LC20.
Thus, the comparison supports more informed exploration by the agent without implying universal superiority over uniform random search.

More broadly, our results show that agentic AI can support density-functional development through the generation, evaluation, and refinement of physically admissible forms.
Our prototype deliberately uses curated term menus, sparse expressions, a fixed gate architecture, and tightly bounded agentic choices, leaving substantial parts of the design space unexplored.
Future workflows can broaden the available analytic forms, allow agents to construct and refine terms beyond a fixed catalog, optimize parameters fixed manually here, and incorporate reduced-gradient behavior relevant to lattice-constant accuracy.

\section*{Acknowledgments}
We thank Philipp T. Ernst and Sam Pallister for critically reviewing the manuscript.
We also thank Arpan Kundu, Stamatia Zavitsanou, and the broader quantum chemistry team at PsiQuantum for helpful discussions and feedback.

\section*{Data Availability}
The \sifull provides the complete set of exchange and correlation terms presented to the LLM.
It also lists the exact constraints, physical norms, residual scales, and weights used to modify and evaluate r2SCAN.
The explicit forms and optimized parameters of the selected candidates are reported, along with the procedures and inputs used for the calculations.
Complete numerical results underlying the molecular and solid-state benchmarks and figures are also provided to support reproducibility.
Further information and code related to this work may be obtained from the corresponding authors upon reasonable request.

\masterbibcontrols
\putbib
\end{bibunit}

\clearpage
\makeatletter
\def\@bibunitname{si_ref}
\makeatother
\begin{bibunit}
\onecolumngrid
\renewcommand{\addcontentsline}[3]{\masteraddcontentsline{#1}{#2}{#3}}
\setcounter{page}{1}
\setcounter{section}{0}
\setcounter{subsection}{0}
\setcounter{figure}{0}
\setcounter{table}{0}
\setcounter{equation}{0}
\renewcommand{\thesection}{S\arabic{section}}
\renewcommand{\thesubsection}{S\arabic{section}.\arabic{subsection}}
\renewcommand{\thefigure}{S\arabic{figure}}
\renewcommand{\thetable}{S\arabic{table}}
\renewcommand{\theequation}{S\arabic{equation}}
\makeatletter
\renewcommand{\p@subsection}{}
\renewcommand{\theHsection}{SI.\arabic{section}}
\renewcommand{\theHsubsection}{SI.\arabic{section}.\arabic{subsection}}
\renewcommand{\theHfigure}{SI.\arabic{figure}}
\renewcommand{\theHtable}{SI.\arabic{table}}
\renewcommand{\theHequation}{SI.\arabic{equation}}
\makeatother
\setcounter{tocdepth}{2}
\setcounter{secnumdepth}{2}
\makeatletter
\@booleantrue\preprintsty@sw
\@booleanfalse\twocolumn@sw
\@booleantrue\titlepage@sw
\input{aps12pt4-2.rtx}
\def\baselinestretch{1.5}
\oddsidemargin=0pt
\evensidemargin=0pt
\let\section\section@preprintsty
\let\subsection\subsection@preprintsty
\let\subsubsection\subsubsection@preprintsty
\renewcommand{\rmdefault}{ptm}
\normalfont
\normalsize
\columnwidth=\textwidth
\linewidth=\textwidth
\hsize=\textwidth
\@colht=\textheight
\@colroom=\textheight
\vsize=\textheight
\makeatother
\mathversion{siptm}
\thinmuskip=2mu
\medmuskip=2.5mu plus 1mu minus 1mu
\thickmuskip=4mu plus 1.5mu minus 1mu
\setlength{\LTcapwidth}{\textwidth}
\let\title\mastertitle
\let\author\masterauthor
\let\affiliation\masteraffiliation
\let\email\masteremail
\let\date\masterdate
\let\maketitle\mastermaketitle
\title[Supplementary Information]{Supplementary Information for
Agentic AI for Density-Functional Development: Revisiting r2SCAN}
\author{Santosh Adhikari}
\affiliation{PsiQuantum, 700 Hansen Way, Palo Alto, California 94304, USA}
\email{sadhikari@psiquantum.com}
\author{Kelsey A. Parker}
\affiliation{PsiQuantum, 700 Hansen Way, Palo Alto, California 94304, USA}
\author{Etinosa Osaro}
\affiliation{PsiQuantum, 700 Hansen Way, Palo Alto, California 94304, USA}
\author{Swagata Roy}
\affiliation{PsiQuantum, 700 Hansen Way, Palo Alto, California 94304, USA}
\author{Dario Rocca}
\affiliation{PsiQuantum, 700 Hansen Way, Palo Alto, California 94304, USA}
\email{drocca@psiquantum.com}
\date{\today}
\let\mastersilabel\label
\renewcommand{\label}[1]{\mastersilabel{SI.#1}}
\maketitle
\let\label\mastersilabel
\onecolumngrid
\clearpage
\thispagestyle{plain}
\pagestyle{plain}
\tableofcontents

\section{Details on candidate construction and physical verification}
\label{sec:agentic-revision}

Here we provide the details of candidate construction and physical verification summarized in Sec.~II\,A of the main manuscript.
We first define the notation and meta-GGA ingredients, then introduce the r2SCAN-based analytic correction family and list the exact exchange and correlation terms presented to the LLM proposer.

\subsection{Notation and terminology for candidate construction}
\label{sec:focused-notation}

We use atomic units throughout.
Let \(n_\uparrow(\mathbf r)\) and \(n_\downarrow(\mathbf r)\) be the spin densities and \(n=n_\uparrow+n_\downarrow\) the total electron density.
The exchange, correlation, and exchange--correlation energies are
\begin{equation}
 \begin{aligned}
  E_x[n_\uparrow,n_\downarrow]
  &=\int n(\mathbf r)\varepsilon_x(\mathbf r)\,d\mathbf r,\\
  E_c[n_\uparrow,n_\downarrow]
  &=\int n(\mathbf r)\varepsilon_c(\mathbf r)\,d\mathbf r,\\
  E_{\mathrm{xc}}&=E_x+E_c
  =\int n(\mathbf r)\varepsilon_{\mathrm{xc}}(\mathbf r)\,d\mathbf r,
  \qquad \varepsilon_{\mathrm{xc}}=\varepsilon_x+\varepsilon_c.
 \end{aligned}
  \label{eq:focused-energy-integral}
\end{equation}
Here, \(\varepsilon_x\), \(\varepsilon_c\), and \(\varepsilon_{\mathrm{xc}}\) are the corresponding energies per electron.

The local meta-GGA ingredients are
\begin{equation}
 \begin{aligned}
  r_s&=\left(\frac{3}{4\pi n}\right)^{1/3},
  &k_F&=(3\pi^2n)^{1/3},\\
  s&=\frac{|\nabla n|}{2k_Fn},
  &p&=s^2,
  &\zeta&=\frac{n_\uparrow-n_\downarrow}{n}.
 \end{aligned}
 \label{eq:focused-semilocal-ingredients}
\end{equation}
Here, \(r_s\) is the Wigner--Seitz radius, \(k_F\) is the local Fermi wave vector, \(s\) is the reduced density gradient, \(p\) is its square, and \(\zeta\) is the relative spin polarization.
The positive Kohn--Sham kinetic-energy density is
\begin{equation}
 \tau=\frac12\sum_{i\sigma}f_{i\sigma}|\nabla\psi_{i\sigma}|^2,
 \qquad
 \tau_W=\frac{|\nabla n|^2}{8n},
 \qquad
 \tau_{\mathrm{unif}}=\frac{3}{10}k_F^2n,
 \label{eq:focused-kinetic-ingredients}
\end{equation}
where \(f_{i\sigma}\) and \(\psi_{i\sigma}\) are orbital occupations and Kohn--Sham orbitals, \(\tau_W\) is the von Weizs\"acker kinetic-energy density, and \(\tau_{\mathrm{unif}}\) is the uniform-gas value.

The iso-orbital input is r2SCAN's regularized indicator~\citep{furness2020r2scan},
\begin{equation}
  \alphareg
  =
  \frac{\tau-\tau_W}
       {\tau_{\mathrm{unif}}+\eta_\alpha\tau_W},
  \qquad
  \eta_\alpha=10^{-3}.
  \label{eq:focused-regularized-alpha}
\end{equation}
Here, \(\eta_\alpha=10^{-3}\) is the fixed r2SCAN regularization parameter.
Throughout the main text and this \sifull, \(\alpha^{\mathrm{r2}}\) denotes this regularized indicator, distinguishing it from the conventional SCAN indicator \(\alpha=(\tau-\tau_W)/\tau_{\mathrm{unif}}\)~\citep{sun2015scan}.

We further define the transformed variables
\begin{equation}
  \bar s=\frac{s}{1+s},
  \qquad
  \bar p=\frac{p}{1+p}=\frac{s^2}{1+s^2},
  \qquad
  \bar\alpha=\frac{\alphareg}{1+\alphareg},
  \qquad
  \bar r_s=\frac{r_s}{1+r_s}.
  \label{eq:focused-bounded-variables}
\end{equation}
In the implementation, negative values of \(\alphareg\) caused by numerical noise are set to zero before these expressions are evaluated.
Thus, \(\bar s\), \(\bar p\), \(\bar\alpha\), and \(\bar r_s\) are bounded forms of the reduced gradient, its square, the regularized iso-orbital indicator, and the Wigner--Seitz radius, respectively.
We use these transformed variables below to construct candidate forms.

\subsection{Candidate construction and analytic term catalogs}
\label{sec:focused-analytic-form}

For candidate \(K\), the exchange and correlation channels are corrected independently from r2SCAN,
\begin{equation}
  \epsx^K=\epsx^{\mathrm{r2SCAN}}+\Delta\epsx^K,
  \qquad
  \epsc^K=\epsc^{\mathrm{r2SCAN}}+\Delta\epsc^K,
  \qquad
  \epsxc^K=\epsx^K+\epsc^K.
  \label{eq:focused-decomposition}
\end{equation}
Here, \(\Delta\epsx^K\) and \(\Delta\epsc^K\) are the candidate corrections to the exchange and correlation energies per electron, respectively.
The two channels are implemented as in Eq.~(3) of the main manuscript,
\begin{equation}
 \begin{aligned}
  \epsx^K&=\epsx^{\mathrm{r2SCAN}}
  +\varepsilon_x^{\mathrm{unif}}
  g_\alpha(\alphareg;\eta_K,\sigma_K)P_x^K,\\
  \epsc^K&=\epsc^{\mathrm{r2SCAN}}
  +\bar r_s
  \left|\epsc^{\mathrm{r2SCAN}}\right|
  g_\alpha(\alphareg;\eta_K,\sigma_K)P_c^K.
 \end{aligned}
 \label{eq:focused-channel-corrections}
\end{equation}
Here, \(g_\alpha\) is the gate introduced in Eq.~(2) of the main manuscript and shared by all selected terms in a proposal; its construction and parameter roles are discussed there.
Its explicit form is
\begin{equation}
  g_\alpha(\alphareg;\eta,\sigma)
  =
  \frac{\alphareg}{\alphareg+\eta}
  \frac{\alphareg-1}
       {1+[(\alphareg-1)/\sigma]^2}.
  \label{eq:focused-alpha-gate}
\end{equation}
Because the gate and every \(\alpha\)-dependent catalog term use \(\alpha^{\mathrm{r2}}\), all candidate forms retain the r2SCAN regularization.
The spin-unpolarized uniform-gas exchange energy per electron is
\begin{equation}
 \varepsilon_x^{\mathrm{unif}}(n)
 =-\frac{3}{4\pi}(3\pi^2n)^{1/3}.
 \label{eq:focused-uniform-exchange}
\end{equation}
The resulting exchange--correlation energy per electron and enhancement factor are
\begin{equation}
  \epsxc^K=\epsx^K+\epsc^K,
  \qquad
  F_{\mathrm{xc}}^K=
  \frac{\epsxc^K}{\varepsilon_x^{\mathrm{unif}}}.
  \label{eq:focused-fxc}
\end{equation}
The quantity \(P_\nu^K\), with \(\nu\in\{x,c\}\), in \eref{eq:focused-channel-corrections} is defined as
\begin{equation}
  P_\nu^K=\sum_{j=1}^{m_{\nu,K}}c_{\nu j}^K\phi_{\nu j},
  \qquad
  1\le m_{x,K}+m_{c,K}\le3.
  \label{eq:focused-candidate-expansion}
\end{equation}
Here, \(m_{\nu,K}\) is the number of selected terms in channel \(\nu\), \(c_{\nu j}^K\) is the corresponding coefficient, and \(\phi_{\nu j}\) is selected from the exchange or correlation catalog presented in Tables~\ref{tab:focused-exchange-menu} and~\ref{tab:focused-correlation-menu}, respectively.
Across the production runs, the proposer selected 54 of the 78 exchange terms and 48 of the 90 correlation terms at least once, as indicated by ``yes'' in the Selected columns of Tables~\ref{tab:focused-exchange-menu} and~\ref{tab:focused-correlation-menu}.

\begingroup
\small
\setlength{\tabcolsep}{6pt}
\renewcommand{\arraystretch}{0.96}
\begin{longtable}{@{}r l c r l c@{}}
\caption{Exact 78-term exchange menu presented to the LLM proposer.}
\label{tab:focused-exchange-menu}\\
\toprule
No. & Exact exchange term & Selected & No. & Exact exchange term & Selected \\
\midrule
\endfirsthead
\multicolumn{6}{c}{\tablename\ \thetable\ (continued)}\\
\toprule
No. & Exact exchange term & Selected & No. & Exact exchange term & Selected \\
\midrule
\endhead
\midrule
\multicolumn{6}{r}{Continued on next page}\\
\endfoot
\bottomrule
\endlastfoot
1 & \code{p} & -- & 40 & \code{s*exp(-s^3)} & yes \\
2 & \code{s} & -- & 41 & \code{s*p*exp(-p)} & yes \\
3 & \code{p^2} & -- & 42 & \code{s*p*exp(-s)} & yes \\
4 & \code{p^3} & -- & 43 & \code{s^2*exp(-p)} & yes \\
5 & \code{s*p} & -- & 44 & \code{s^2*exp(-s)} & yes \\
6 & \code{s^2} & -- & 45 & \code{exp(-p)} & yes \\
7 & \code{s^3} & -- & 46 & \code{exp(-s)} & -- \\
8 & \code{alpha} & -- & 47 & \code{exp(-p^2)} & -- \\
9 & \code{s*p^2} & -- & 48 & \code{exp(-p^3)} & yes \\
10 & \code{s^2*p} & -- & 49 & \code{exp(-s^2)} & yes \\
11 & \code{alpha^2} & -- & 50 & \code{exp(-s^3)} & -- \\
12 & \code{alpha^3} & -- & 51 & \code{exp(-alpha)} & yes \\
13 & \code{p*alpha} & yes & 52 & \code{exp(-alpha^2)} & yes \\
14 & \code{s*alpha} & yes & 53 & \code{exp(-alpha^3)} & yes \\
15 & \code{p*alpha^2} & yes & 54 & \code{p/(1+p)} & yes \\
16 & \code{p^2*alpha} & yes & 55 & \code{p/(1+s)} & yes \\
17 & \code{s*alpha^2} & yes & 56 & \code{s/(1+p)} & -- \\
18 & \code{s*p*alpha} & yes & 57 & \code{s/(1+s)} & -- \\
19 & \code{s^2*alpha} & yes & 58 & \code{p/(1+p^2)} & yes \\
20 & \code{exp(-s*p)} & yes & 59 & \code{p/(1+p^3)} & yes \\
21 & \code{p*exp(-p)} & yes & 60 & \code{p/(1+s*p)} & yes \\
22 & \code{p*exp(-s)} & yes & 61 & \code{p/(1+s^2)} & yes \\
23 & \code{s*exp(-p)} & yes & 62 & \code{p/(1+s^3)} & -- \\
24 & \code{s*exp(-s)} & yes & 63 & \code{p^2/(1+p)} & yes \\
25 & \code{exp(-s*p^2)} & yes & 64 & \code{p^2/(1+s)} & yes \\
26 & \code{exp(-s^2*p)} & yes & 65 & \code{p^3/(1+p)} & -- \\
27 & \code{p*exp(-p^2)} & yes & 66 & \code{p^3/(1+s)} & yes \\
28 & \code{p*exp(-p^3)} & yes & 67 & \code{s*p/(1+p)} & yes \\
29 & \code{p*exp(-s*p)} & yes & 68 & \code{s*p/(1+s)} & yes \\
30 & \code{p*exp(-s^2)} & yes & 69 & \code{s/(1+p^2)} & yes \\
31 & \code{p*exp(-s^3)} & yes & 70 & \code{s/(1+p^3)} & -- \\
32 & \code{p^2*exp(-p)} & yes & 71 & \code{s/(1+s*p)} & -- \\
33 & \code{p^2*exp(-s)} & yes & 72 & \code{s/(1+s^2)} & yes \\
34 & \code{p^3*exp(-p)} & yes & 73 & \code{s/(1+s^3)} & -- \\
35 & \code{p^3*exp(-s)} & yes & 74 & \code{s^2/(1+p)} & yes \\
36 & \code{s*exp(-p^2)} & yes & 75 & \code{s^2/(1+s)} & yes \\
37 & \code{s*exp(-p^3)} & yes & 76 & \code{s^3/(1+p)} & -- \\
38 & \code{s*exp(-s*p)} & -- & 77 & \code{s^3/(1+s)} & yes \\
39 & \code{s*exp(-s^2)} & yes & 78 & \code{alpha/(1+p)} & yes \\
\end{longtable}
\endgroup

\begingroup
\small
\setlength{\tabcolsep}{6pt}
\renewcommand{\arraystretch}{0.94}
\begin{longtable}{@{}r l c r l c@{}}
\caption{Exact 90-term correlation menu presented to the LLM proposer.}
\label{tab:focused-correlation-menu}\\
\toprule
No. & Exact correlation term & Selected & No. & Exact correlation term & Selected \\
\midrule
\endfirsthead
\multicolumn{6}{c}{\tablename\ \thetable\ (continued)}\\
\toprule
No. & Exact correlation term & Selected & No. & Exact correlation term & Selected \\
\midrule
\endhead
\midrule
\multicolumn{6}{r}{Continued on next page}\\
\endfoot
\bottomrule
\endlastfoot
1 & \code{p} & -- & 46 & \code{p^3*exp(-p)} & -- \\
2 & \code{s} & -- & 47 & \code{p^3*exp(-s)} & -- \\
3 & \code{rs} & -- & 48 & \code{rs*exp(-rs)} & yes \\
4 & \code{p^2} & -- & 49 & \code{s*exp(-p^2)} & yes \\
5 & \code{p^3} & -- & 50 & \code{s*exp(-p^3)} & -- \\
6 & \code{s*p} & -- & 51 & \code{exp(-p)} & -- \\
7 & \code{s^2} & -- & 52 & \code{exp(-s)} & -- \\
8 & \code{s^3} & -- & 53 & \code{exp(-rs)} & yes \\
9 & \code{rs*p} & -- & 54 & \code{exp(-p^2)} & -- \\
10 & \code{rs*s} & -- & 55 & \code{exp(-p^3)} & -- \\
11 & \code{rs^2} & -- & 56 & \code{exp(-s^2)} & -- \\
12 & \code{rs^3} & -- & 57 & \code{exp(-s^3)} & -- \\
13 & \code{alpha} & yes & 58 & \code{exp(-rs^2)} & yes \\
14 & \code{s*p^2} & -- & 59 & \code{exp(-rs^3)} & yes \\
15 & \code{s^2*p} & -- & 60 & \code{exp(-alpha)} & yes \\
16 & \code{zeta2} & -- & 61 & \code{exp(-zeta2)} & yes \\
17 & \code{rs*p^2} & -- & 62 & \code{exp(-alpha^2)} & yes \\
18 & \code{rs*s*p} & yes & 63 & \code{exp(-alpha^3)} & yes \\
19 & \code{rs*s^2} & -- & 64 & \code{exp(-zeta2^2)} & yes \\
20 & \code{rs^2*p} & yes & 65 & \code{exp(-zeta2^3)} & yes \\
21 & \code{rs^2*s} & yes & 66 & \code{p/(1+p)} & yes \\
22 & \code{alpha^2} & yes & 67 & \code{p/(1+s)} & -- \\
23 & \code{alpha^3} & yes & 68 & \code{s/(1+p)} & yes \\
24 & \code{p*alpha} & yes & 69 & \code{s/(1+s)} & yes \\
25 & \code{p*zeta2} & yes & 70 & \code{p/(1+rs)} & yes \\
26 & \code{exp(-s*p)} & -- & 71 & \code{rs/(1+p)} & yes \\
27 & \code{p*exp(-p)} & yes & 72 & \code{rs/(1+s)} & yes \\
28 & \code{p*exp(-s)} & yes & 73 & \code{s/(1+rs)} & yes \\
29 & \code{s*exp(-p)} & yes & 74 & \code{p/(1+p^2)} & yes \\
30 & \code{s*exp(-s)} & yes & 75 & \code{p/(1+p^3)} & yes \\
31 & \code{exp(-rs*p)} & -- & 76 & \code{p/(1+s*p)} & yes \\
32 & \code{exp(-rs*s)} & -- & 77 & \code{p/(1+s^2)} & yes \\
33 & \code{p*exp(-rs)} & yes & 78 & \code{p/(1+s^3)} & yes \\
34 & \code{rs*exp(-p)} & yes & 79 & \code{p^2/(1+p)} & -- \\
35 & \code{rs*exp(-s)} & yes & 80 & \code{p^2/(1+s)} & yes \\
36 & \code{s*exp(-rs)} & yes & 81 & \code{p^3/(1+p)} & -- \\
37 & \code{exp(-s*p^2)} & -- & 82 & \code{p^3/(1+s)} & -- \\
38 & \code{exp(-s^2*p)} & -- & 83 & \code{rs/(1+rs)} & yes \\
39 & \code{p*exp(-p^2)} & -- & 84 & \code{s*p/(1+p)} & yes \\
40 & \code{p*exp(-p^3)} & -- & 85 & \code{s*p/(1+s)} & yes \\
41 & \code{p*exp(-s*p)} & -- & 86 & \code{s/(1+p^2)} & yes \\
42 & \code{p*exp(-s^2)} & yes & 87 & \code{s/(1+p^3)} & -- \\
43 & \code{p*exp(-s^3)} & -- & 88 & \code{s/(1+s*p)} & -- \\
44 & \code{p^2*exp(-p)} & yes & 89 & \code{s/(1+s^2)} & yes \\
45 & \code{p^2*exp(-s)} & yes & 90 & \code{s/(1+s^3)} & -- \\
\end{longtable}
\endgroup

Thus, a candidate contains one to three terms in total across either or both channels.
For spin-polarized densities, the exchange channel is applied through exact spin scaling, while the correlation channel uses the total-density ingredients and may depend explicitly on \(\zeta^2\).
Setting all coefficients to zero recovers r2SCAN exactly.

\subsection{Constraint checks for candidates}
\label{sec:focused-constraints}

The agentic workflow requires every accepted correction to retain the 16 exact constraints satisfied by r2SCAN.
Table~\ref{tab:focused-constraints} groups these conditions into exchange, correlation, and combined exchange--correlation blocks and summarizes how each is preserved.
Here, ``by construction'' denotes a consequence of the analytic correction form, whereas ``checked'' denotes a deterministic numerical test over the prescribed domains and tolerances.
The complete fourth-order exchange gradient expansion (GE4X) is omitted because r2SCAN does not satisfy that SCAN constraint~\citep{furness2020r2scan}.
At the proposal-and-critique stage, the workflow uses inexpensive, tolerance-based checks on finite grids of synthetic density ingredients to screen the initial form and coefficient values.
During coefficient optimization, it evaluates four scaling relations more tightly as continuous normalized residuals that enter candidate scoring: exchange uniform scaling, exchange nonuniform scaling, correlation high-density uniform scaling, and correlation low-density uniform scaling.

\begingroup
\footnotesize
\renewcommand{\arraystretch}{1.04}
\newcommand{\constraintname}[1]{\parbox[t]{5.8cm}{\raggedright #1\par}}
\newcommand{\constraintstatus}[1]{\parbox[t]{9.0cm}{\raggedright #1\par}}
\begin{longtable}{ll}
\caption{List of the 16 exact constraints retained by the candidates and a brief description of how they are treated in the candidate workflow.}
\label{tab:focused-constraints}\\
\toprule
\constraintname{Constraint retained by r2SCAN} & \constraintstatus{Treatment in the candidate workflow} \\
\midrule
\endfirsthead
\multicolumn{2}{c}{\tablename~\thetable\ (continued)}\\
\toprule
\constraintname{Constraint retained by r2SCAN} & \constraintstatus{Treatment in the candidate workflow} \\
\midrule
\endhead
\midrule
\multicolumn{2}{r}{Continued on next page}\\
\endfoot
\bottomrule
\endlastfoot
\multicolumn{2}{l}{\textbf{Exchange}} \\
\constraintname{Nonpositivity of exchange} & \constraintstatus{Checked by requiring \(\varepsilon_x^K\le 0\) over a synthetic grid spanning \((r_s,s,\alphareg,\zeta)\).} \\
\constraintname{Exact exchange spin scaling} & \constraintstatus{Satisfied by construction through exact spin scaling of the exchange channel.} \\
\constraintname{Uniform coordinate scaling of exchange} & \constraintstatus{Satisfied by the dimensionless exchange construction and checked along synthetic uniform-density scaling paths.} \\
\constraintname{Nonuniform coordinate scaling of exchange} & \constraintstatus{Checked along synthetic nonuniform-scaling paths that approach the large-gradient limit.} \\
\constraintname{Tight exchange bound for two-electron densities} & \constraintstatus{Checked by requiring \(F_x^K\le 1.174\) for synthetic one-orbital and two-electron densities.} \\
\addlinespace
\multicolumn{2}{l}{\textbf{Correlation}} \\
\constraintname{Nonpositivity of correlation} & \constraintstatus{Checked by requiring \(\varepsilon_c^K\le 0\) over a synthetic grid spanning \((r_s,s,\alphareg,\zeta)\).} \\
\constraintname{Second-order correlation gradient expansion (GE2C)} & \constraintstatus{Checked near the spin-unpolarized, high-density, slowly varying limit against \(\beta_0=0.0667245506\).} \\
\constraintname{High-density uniform scaling of correlation} & \constraintstatus{Checked along synthetic paths that increase the density while holding the dimensionless ingredients fixed.} \\
\constraintname{Low-density uniform scaling of correlation} & \constraintstatus{Checked along synthetic paths approaching the low-density limit.} \\
\constraintname{Zero correlation energy for one-electron densities} & \constraintstatus{Satisfied by construction because the gate vanishes at \(\alphareg=0\), recovering r2SCAN.} \\
\constraintname{Nonuniform coordinate scaling of correlation} & \constraintstatus{Checked using a synthetic two-electron density compressed along one coordinate.} \\
\addlinespace
\multicolumn{2}{l}{\textbf{Exchange--correlation}} \\
\constraintname{Size extensivity} & \constraintstatus{Satisfied by construction through the semilocal energy integral.} \\
\constraintname{General Lieb--Oxford bound} & \constraintstatus{Checked by integrating analytic one- and two-electron model densities and requiring \(E_{\mathrm{xc}}/E_x^{\mathrm{LDA}}\le 2.215\).} \\
\constraintname{Weak spin-polarization dependence in the low-density limit} & \constraintstatus{Checked over a synthetic grid of large \(r_s\), reduced gradients, and spin polarizations.} \\
\constraintname{Uniform-electron-gas static linear response} & \constraintstatus{Checked from the combined second-order exchange--correlation response to a weak synthetic density modulation.} \\
\constraintname{Tight two-electron Lieb--Oxford bound} & \constraintstatus{Checked by integrating analytic two-electron model densities and requiring \(E_{\mathrm{xc}}/E_x^{\mathrm{LDA}}\le 1.67082\).} \\
\end{longtable}
\endgroup

The verifier additionally checks that the second-order exchange gradient coefficient remains \(\mu=10/81\).
For r2SCAN, this retained second-order coefficient is enforced while the fourth-order exchange-gradient condition is not imposed, because r2SCAN does not satisfy the latter.

\subsection{Nonempirical norms used for candidate verification}
\label{sec:focused-norm-inventory}

After a candidate passes the constraint checks, the workflow evaluates six nonempirical norm categories.
No molecular reaction energy, equilibrium bond length, lattice constant, or band gap enters any of these categories.
We did not include jellium surface formation energies as a separate candidate-level norm in this study, although they were used as appropriate norms in constructing r2SCAN~\citep{furness2020r2scan}.

\begin{enumerate}[leftmargin=*,itemsep=0.45em]
\item \textit{Scaling.}
Coordinate-scaling relations are exact constraints enforced by the deterministic verifier described in \sref{sec:focused-constraints}~\citep{sun2015scan,furness2020r2scan}.
Because their asymptotic limits are represented numerically by finite scaling factors, grids, and large-gradient windows, here we use the resulting residuals to measure how far each candidate drifts from the identically evaluated r2SCAN behavior at the sampled scales.
The four scored relations are exchange uniform scaling, measured from the error in \(\varepsilon_x[n_\gamma]/\varepsilon_x[n]=\gamma\) at \(\gamma=2\); exchange nonuniform scaling, measured from the large-\(s\) slope of \(\log F_x\) relative to \(-1/2\); correlation high-density scaling, measured from the terminal slope of \(\varepsilon_c\) relative to zero; and correlation low-density scaling, measured from the final relative change in \(\varepsilon_c/\gamma\) relative to zero.
Their grids and normalization are given in \sref{sec:focused-norm-weights}; correlation nonuniform scaling remains a pass--fail verifier check and is not included in this continuous norm.

\item \textit{Rare-gas exchange.}
Candidate exchange energies are integrated on fixed restricted-Hartree--Fock densities using the PySCF level-6 grid.
Ne, Ar, and Kr use def2-QZVPPD, whereas Xe uses a fully uncontracted all-electron ANO-RCC basis.
The reference total exchange energies are \((-12.1050,-30.1752,-93.8340,-179.0640)\) Ha, obtained from the per-electron exact-exchange values tabulated by Burke \textit{et al.}~\citep{burke2014atomiccorrelation}.

\item \textit{Rare-gas correlation.}
Candidate correlation energies are integrated on the same fixed rare-gas Hartree--Fock densities.
The corresponding benchmark reference energies are \((-0.3910,-0.7254,-1.8504,-3.0024)\) Ha for Ne, Ar, Kr, and Xe, respectively~\citep{burke2014atomiccorrelation}.

\item \textit{Large-\(Z\) coefficients.}
Using the candidate exchange and correlation energies for the same four atoms, the workflow fits
\[
 E_x-E_x^{\mathrm{LDA}}=\gamma_{x1}Z+\gamma_{x2}Z^{2/3},
 \qquad
 E_c-E_c^{\mathrm{LDA}}=\gamma_{c1}Z.
\]
The fitted coefficients are properties of candidate \(K\).
As fixed anchors, we use the SCAN-reported values~\citep{sun2015scan}
\[
 (\gamma_{x1},\gamma_{x2},\gamma_{c1})
 =(-0.2259,0.2551,0.0388).
\]
Both r2SCAN and every candidate are compared with the same anchors.
The resulting normalized residuals enter the aggregate norm loss used to decide whether a candidate remains within the r2SCAN loss budget.

\item \textit{Compressed Ar\(_2\).}
At separations \(R=(3.023562,3.401507,3.779452)\,a_0\), the workflow evaluates candidate interaction energies non-self-consistently on fixed r2SCAN densities and forms their mean absolute error.
The reference interaction energies from the high-accuracy Ar\(_2\) potential of Patkowski \textit{et al.}~\citep{patkowski2005argon} are
\[
 (0.5821383510,0.3159435766,0.1644855252)\ \mathrm{Ha}.
\]

\item \textit{Fractional-charge linearity.}
For a fixed hydrogenic \(1s\), \(Z=1\) ensemble, the workflow first evaluates \(E_H+E_{\mathrm{xc}}\) at the integer occupations \(N=0,1,2\).
These three endpoint values define one straight-line reference over \(0\le N\le 1\) and a second over \(1\le N\le 2\).
The workflow then evaluates the same quantity at \(N=0.25,0.50,0.75,1.25,1.50,1.75\) and scores its departures from the corresponding interpolated reference.
All evaluations use an 8000-point radial grid over \([10^{-5},25]a_0\), so this category measures piecewise linearity within an internally defined fixed-potential ensemble rather than against an external bonded-system reference.
\end{enumerate}

\subsection{Residual scales, grids, and weights}
\label{sec:focused-norm-weights}

For the inexpensive coefficient search, the scaling loss \(\mathcal L_{\gamma}^{K}\) in Eq.~(6) of the main manuscript is
\begin{equation}
 \mathcal L_{\gamma}^{K}
 =\left[
 \frac{1}{4}\sum_{a=1}^{4}
 \left(\frac{P_a^K-R_a}{s_a}\right)^2
 \right]^{1/2}.
 \label{eq:focused-scaling-loss}
\end{equation}
Here, \(P_a^K\) is the finite-grid residual for scaling relation \(a\), every exact target is \(R_a=0\), and the four relations have equal weight.
Their fixed normalization scales are
\begin{equation}
 \begin{array}{c|cccc}
 a & \text{exchange uniform} & \text{exchange nonuniform}
   & \text{correlation high density} & \text{correlation low density}\\
 \hline
 s_a & 10^{-5} & 0.05 & 10^{-3} & 0.12.
 \end{array}
 \label{eq:focused-scaling-scales}
\end{equation}
We prescribed these scales manually before the search from the characteristic numerical magnitudes of the four diagnostics, placing them on comparable dimensionless footing; they were not optimized.
We obtain \(\mathcal L_{\gamma}^{\mathrm{r2SCAN}}\) from the same equation, grids, and scales.

The second auxiliary quantity in Eq.~(6) of the main manuscript is
\begin{equation}
 \mathcal L_{\Delta F_{\mathrm{xc}}}^{K}
 =\left[
 \frac{1}{30}\sum_{r_s}\sum_s
 \left(F_{\mathrm{xc}}^K-F_{\mathrm{xc}}^{\mathrm{r2SCAN}}\right)^2
 \right]^{1/2},
 \label{eq:focused-fxc-departure}
\end{equation}
with equal point weights on \(r_s=(2,3,4,5,6)\) and \(s=(0,0.2,0.5,0.8,1.2,1.5)\) at \(\zeta=0\) and \(\alpha^{\mathrm{r2}}=[1+(5/3)10^{-3}s^2]^{-1}\).

The derivative-response ratio \(R_{\alpha^{\mathrm{r2}}}^{K}\) uses equal point weights at \(\zeta=0\) on
\[
 r_s=(0.5,1,2,5),\qquad s=(0,0.2,0.5,1,2),\qquad
 a=(0.5,1,2,3,5,10).
\]
The regularized-indicator samples are \(\alpha^{\mathrm{r2}}=a/[1+(5/3)10^{-3}s^2]\), and centered differences use a step of \(10^{-4}\) in \(a\).

\begin{figure}[t]
 \centering
 \includegraphics[width=0.94\linewidth]{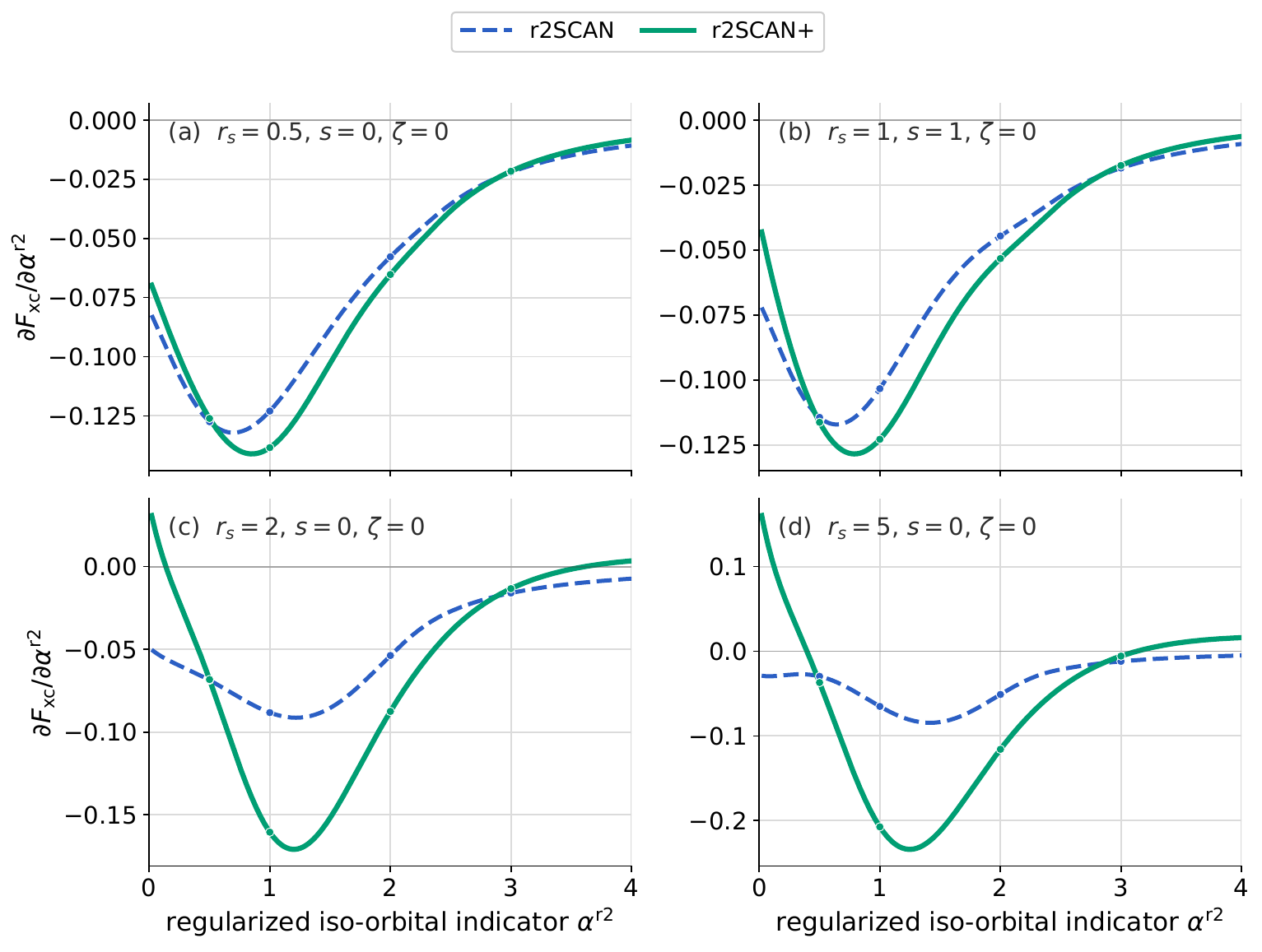}
\caption{Representative \((r_s,s)\) slices of the response of \(F_{\mathrm{xc}}\) to the regularized iso-orbital indicator for r2SCAN and r2SCAN+.
 These slices complement the main-text slice used in Fig.~1 and sample distinct density and reduced-gradient regimes from the diagnostic space.}
 \label{fig:focused-alpha-slices}
\end{figure}

Finally, the full-verification loss \(\mathcal L_{\mathrm{norm}}\) in Eq.~(8) of the main manuscript uses the fixed residual scales \(s_i\) and category weights \(W_g=1/M_g^{\mathrm{r2SCAN}}\) in \tref{tab:focused-norm-weights}; every datum has weight \(w_i=1\).
The residual scales \(s_i\) convert deviations of different magnitudes into comparable dimensionless quantities before they are averaged within each category.
For the atomic energies and large-\(Z\) coefficients, we set these scales from the magnitudes of the corresponding reference values; the other categories use fixed characteristic scales associated with their diagnostics.
The category weights balance the six category mean squares so that they contribute equally at the r2SCAN baseline.
These manually prescribed choices provide a simple and convenient normalization, but we do not claim that the scales or category weights are optimal.

\begin{table}[H]
\centering
\scriptsize
\setlength{\tabcolsep}{3pt}
\renewcommand{\arraystretch}{1.12}
\newcommand{\normcategory}[1]{\parbox[t]{1.00in}{\raggedright #1\par}}
\newcommand{\normquantity}[1]{\parbox[t]{1.20in}{\raggedright #1\par}}
\newcommand{\normscale}[1]{\parbox[t]{2.55in}{\raggedright #1\par}}
\newcommand{\normweight}[1]{\parbox[t]{1.30in}{\raggedleft #1\par}}
\caption{Residual scales and category weights used in the full physical-norm loss.
Residual scales are listed in the quantity order specified in each row; all point weights are \(w_i=1\).}
\label{tab:focused-norm-weights}
\begin{tabular}{@{}llll@{}}
\toprule
\normcategory{Category \(g\)} & \normquantity{Quantity order} & \normscale{Residual scales \(s_i\)} & \normweight{Category weight \(W_g\)} \\
\midrule
\normcategory{Scaling} & \normquantity{Four relations in \eref{eq:focused-scaling-scales}}
  & \normscale{Values in \eref{eq:focused-scaling-scales}}
  & \normweight{\(1.381132592333317\!\times\!10^{-3}\)} \\
\normcategory{Rare-gas exchange} & \normquantity{Ne, Ar, Kr, Xe}
  & \normscale{\((12.1050,\ 30.1752,\ 93.8340,\ 179.0640)\)}
  & \normweight{\(3.295969653466588\!\times\!10^{5}\)} \\
\normcategory{Rare-gas correlation} & \normquantity{Ne, Ar, Kr, Xe}
  & \normscale{\((0.3910,\ 0.7254,\ 1.8504,\ 3.0024)\)}
  & \normweight{\(2.377386485714671\!\times\!10^{2}\)} \\
\normcategory{Large \(Z\)} & \normquantity{\(\gamma_{x1},\gamma_{x2},\gamma_{c1}\)}
  & \normscale{\((0.2259,\ 0.2551,\ 0.0388)\)}
  & \normweight{\(7.198601861425271\!\times\!10^{1}\)} \\
\normcategory{Compressed Ar\(_2\)} & \normquantity{Three-point MAE}
  & \normscale{\(0.001593601437641\)}
  & \normweight{\(1.144459771386029\!\times\!10^{-1}\)} \\
\normcategory{Fractional charge} & \normquantity{Nine linearity residuals}
  & \normscale{\(0.5898124339\)}
  & \normweight{\(3.156819304122731\!\times\!10^{2}\)} \\
\bottomrule
\end{tabular}
\end{table}
This normalization makes the six categories contribute equally at the r2SCAN baseline, \(\mathcal L_{\mathrm{norm}}^{\mathrm{r2SCAN}}=0.004262577643\).
Applying this normalization retains 1,150 of the 1,399 proposals that reached full physical verification, including the five nondominated candidates from the agentic search.

\section{Uniform random search implementation details and complete search statistics}
\label{sec:focused-random-comparison}

The uniform random search used the same curated catalogs, fixed gate architecture, objective, corrected norm protocol, and deterministic verifier as the agentic workflow, with randomized proposal selection.
Across all initializations, the uniform random search generated 2,480 proposals.
This section provides the implementation details, complete search statistics, and the highest-response candidates; the main text discusses the interpretation.

\subsection{Top-response candidate lists}
\label{sec:top-response-candidates}

Tables~\ref{tab:top-agentic-candidates} and~\ref{tab:top-manual-candidates} list the ten highest-response candidates in each search arm, ranked by the verified derivative-response opening among candidates within the r2SCAN norm-loss budget.
The first three entries in the agentic list are r2SCAN+, Finalist~2, and Finalist~3, whose analytic forms are given in Eq.~(9) of the main text.

\begin{table}[H]
\centering
\caption{Ten highest-response agentic candidates.
The first three are the finalists reported in Eq.~(9) of the main text.
Coefficients are the final optimized values multiplying the displayed basis terms.}\label{tab:top-agentic-candidates}
\begin{tabular}{@{}lll@{}}
\toprule
Rank & Explicit form with optimized coefficients & $R_{\alpha^{\mathrm{r2}}}$ \\
\midrule
1 & $0.236802e^{-\bar\alpha}+0.390769e^{-\bar\alpha^2}+0.498078e^{-\bar\alpha^3}$ (r2SCAN+) & 1.3305 \\
2 & $0.485686e^{-\bar\alpha^2}+0.495912e^{-\bar\alpha^3}+0.193335\bar r_s e^{-\bar r_s}$ (Finalist~2) & 1.3287 \\
3 & $0.499610e^{-\bar\alpha}+0.499965e^{-\bar\alpha^3}+0.199782\bar r_s/(1+\bar s)$ (Finalist~3) & 1.3282 \\
4 & $0.497308e^{-\bar\alpha^2}+0.496218e^{-\bar\alpha^3}+0.108661\bar\alpha^3$ & 1.3127 \\
5 & $0.596974e^{-\bar\alpha}+0.597699e^{-\bar\alpha^2}+0.015050\bar p e^{-\bar s^2}$ & 1.3114 \\
6 & $0.499963e^{-\bar\alpha^2}+0.493624e^{-\bar\alpha^3}+0.472644\bar p^3/(1+\bar s)$ & 1.3107 \\
7 & $0.499963e^{-\bar\alpha^2}+0.493624e^{-\bar\alpha^3}+0.283587\bar p^2/(1+\bar p)$ & 1.3100 \\
8 & $0.498134e^{-\bar\alpha^2}+0.499929e^{-\bar\alpha^3}+0.099386\bar p e^{-\bar s}$ & 1.3087 \\
9 & $0.180132\bar p^2e^{-\bar s}+0.499049e^{-\bar\alpha^2}+0.498102e^{-\bar\alpha^3}$ & 1.3076 \\
10 & $0.593078e^{-\bar\alpha}+0.598177e^{-\bar\alpha^2}-0.065861\bar p e^{-\bar p^2}$ & 1.3049 \\
\bottomrule
\end{tabular}
\end{table}

\begin{table}[H]
\centering
\caption{Ten highest-response candidates from the uniform random search.
None contains an explicit $\alpha$-dependent term; coefficients are the optimized values multiplying the displayed basis terms.}\label{tab:top-manual-candidates}
\begin{tabular}{@{}lll@{}}
\toprule
Rank & Explicit form with optimized coefficients & $R_{\alpha^{\mathrm{r2}}}$ \\
\midrule
1 & $-0.030318\bar s^2/(1+\bar s)+0.459912e^{-\bar r_s\bar p}+0.399523\bar r_s e^{-\bar s}$ (Random-Search Best) & 1.2627 \\
2 & $0.443706e^{-\bar s}+0.054185\bar p^3e^{-\bar s}+0.491249\bar r_s e^{-\bar p}$ (Random-Search~II) & 1.2611 \\
3 & $0.437086e^{-\bar p}+0.267792e^{-\bar s\bar p^2}-0.007524\bar s/(1+\bar s\bar p)$ (Random-Search~III) & 1.2412 \\
4 & $0.374032e^{-\bar r_s\bar p}+0.221813\bar p e^{-\bar p^2}+0.393360\bar r_s e^{-\bar p}$ & 1.2340 \\
5 & $0.005331\bar p/(1+\bar p^2)+0.241997e^{-\bar s\bar p}+0.434651e^{-\bar s^2}$ & 1.2331 \\
6 & $0.368095e^{-\bar r_s\bar s}+0.292589e^{-\bar s^2\bar p}+0.191446\bar p^3/(1+\bar s)$ & 1.2292 \\
7 & $0.318265e^{-\bar p}+0.378716e^{-\bar s\bar p^2}-0.298330\bar p e^{-\bar s}$ & 1.2285 \\
8 & $0.430335e^{-\zeta^2}+0.189431\bar p/(1+\bar s\bar p)+0.184077\bar r_s/(1+\bar p)$ & 1.2213 \\
9 & $0.413311e^{-\bar s\bar p^2}+0.403772\bar r_s/(1+\bar r_s)+0.070877\bar s/(1+\bar r_s)$ & 1.2186 \\
10 & $0.426349e^{-\bar p}-0.093035\bar p^3e^{-\bar p}+0.377279\bar r_s^2$ & 1.2180 \\
\bottomrule
\end{tabular}
\end{table}

\begin{table}[H]
\centering
\scriptsize
\setlength{\tabcolsep}{4pt}
\renewcommand{\arraystretch}{1.1}
\caption{Comparison between the agentic search and the uniform random search.
Percentages are evaluated over the complete funnel for the agentic search and over all proposals of the uniform random search.
Failed precheck denotes structural infeasibility after the candidate reached the optimizer; NA means not applicable because the random search had no critic.}
\label{tab:focused-random-headtohead}
\begin{tabular}{@{}lrrrrrrr@{}}
\toprule
Search & Proposals & Rejected by critic & Failed Precheck & In budget & Exchange-only & With \(\alpha^{\mathrm{r2}}\) & Best \(R_{\alpha^{\mathrm{r2}}}\) \\
 & & & (\%) & (\%) & (\%) & (\%) & \\
\midrule
Agentic & 1738 & 330 & 0.64 & 66.17 & 12.72 & 70.71 & 1.3305 \\
Random-Search & 2480 & NA & 24.56 & 51.21 & 9.76 & 32.58 & 1.2627 \\
\bottomrule
\end{tabular}
\end{table}

\section{Benchmark results}
\label{sec:focused-results}

We implemented each closed-form candidate independently in PySCF~\citep{sun2018pyscf,sun2020pyscf} and Quantum ESPRESSO~\citep{giannozzi2009,giannozzi2017,giannozzi2020}.
The computational details are given in Sec.~III (Computational Details) of the main manuscript.
Here we report r2SCAN+, Finalist~2, Finalist~3, and Random-Search Best, the highest-band-gap-response candidate from the random search, alongside r2SCAN, SCAN, and PBE for most properties.
BH6 contains six representative reactions from DBH22 and is reported separately as a compact benchmark.
Aggregate barrier-height ME and MAE values use only the 22 DBH22 and 46 BH46 reactions, for 68 unique reactions in total, thereby avoiding double counting.

\subsection{Grid convergence of the molecular benchmarks}
\begingroup
\renewcommand{\theHtable}{grid}
\scriptsize
\setlength{\tabcolsep}{3.0pt}
\renewcommand{\arraystretch}{0.96}
\begin{longtable}{@{}llrrrrrrrrr@{}}
\caption*{Mean absolute errors, in kcal/mol, across PySCF grid levels 2--6.
The W4-nMR, W4-MR, BH6, DBH22, BH46, G21IP, and G21EA subsets contain 183, 17, 6, 22, 46, 36, and 25 reactions, respectively; All contains 329 unique reactions and excludes the BH6 duplication.}\\
\toprule
Method & Grid & W4-nMR & W4-MR & BH6 & DBH22 & BH46 & G21IP & G21EA & All \\
\midrule
\endfirsthead
\multicolumn{11}{c}{Grid-convergence results (continued)}\\
\toprule
Method & Grid & W4-nMR & W4-MR & BH6 & DBH22 & BH46 & G21IP & G21EA & All \\
\midrule
\endhead
\midrule
\multicolumn{11}{r}{Continued on next page}\\
\endfoot
\bottomrule
\endlastfoot
PBE & 2 & 17.627 & 29.463 & 9.230 & 8.123 & 9.342 & 3.820 & 3.178 & 13.836 \\
PBE & 3 & 17.627 & 29.463 & 9.230 & 8.123 & 9.342 & 3.820 & 3.178 & 13.836 \\
PBE & 4 & 17.627 & 29.464 & 9.230 & 8.123 & 9.343 & 3.820 & 3.178 & 13.836 \\
PBE & 5 & 17.626 & 29.463 & 9.230 & 8.123 & 9.343 & 3.820 & 3.178 & 13.836 \\
PBE & 6 & 17.627 & 29.464 & 9.230 & 8.123 & 9.343 & 3.820 & 3.178 & 13.836 \\
\midrule
B3LYP & 2 & 4.944 & 6.188 & 4.596 & 3.949 & 4.494 & 3.739 & 2.466 & 4.559 \\
B3LYP & 3 & 4.944 & 6.185 & 4.595 & 3.949 & 4.494 & 3.739 & 2.466 & 4.558 \\
B3LYP & 4 & 4.944 & 6.185 & 4.595 & 3.949 & 4.495 & 3.739 & 2.466 & 4.558 \\
\end{longtable}
\addtocounter{table}{-1}
\endgroup

\subsection{Detailed benchmark values}

The detailed molecular values in Table~S8 use grid level 6 for every method.
Throughout this subsection, Random-Search Best is represented as RS Best.
\begingroup
\scriptsize
\setlength{\tabcolsep}{2.3pt}
\renewcommand{\arraystretch}{0.92}
\begin{longtable}{@{}lrrrrrrrrr@{}}
\caption{Complete values for the 329-reaction GSCDB137 validation suite~\citep{liang2025gscdb137,liang2025gscdbgithub}, comprising 200 W4-17 atomization energies, 68 unique barrier heights, 36 ionization potentials, and 25 electron affinities.
The six-reaction BH6 subset is also tabulated separately but excluded from aggregate values.
Values are in kcal/mol, reaction labels reproduce the GSCDB137 identifiers, and all calculations use the PySCF level-6 grid.
ME is calculated minus reference; statistics use unrounded values.
Random-Search Best is represented as RS Best.}\label{tab:focused-molecular-values}\\
\toprule
Reaction & PBE & B3LYP & SCAN & r2SCAN & r2SCAN+ & \shortstack{Finalist\\2} & \shortstack{Finalist\\3} & RS Best & Reference \\
\midrule
\endfirsthead
\multicolumn{10}{c}{\tablename\ \thetable\ (continued)}\\
\toprule
Reaction & PBE & B3LYP & SCAN & r2SCAN & r2SCAN+ & \shortstack{Finalist\\2} & \shortstack{Finalist\\3} & RS Best & Reference \\
\midrule
\endhead
\midrule
\multicolumn{10}{r}{Continued on next page}\\
\endfoot
\bottomrule
\endlastfoot
\multicolumn{10}{l}{\textit{Atomization energies}} \\
\multicolumn{10}{l}{\textit{TAE-W4-17-nonMR (183)}} \\
\texttt{TAE\_W4-17\_1} & 694.28 & 675.61 & 679.08 & 678.86 & 676.82 & 676.71 & 676.72 & 676.73 & 677.85 \\
\texttt{TAE\_W4-17\_2} & 827.70 & 798.42 & 806.96 & 806.95 & 803.07 & 802.97 & 802.98 & 803.35 & 803.97 \\
\texttt{TAE\_W4-17\_3} & 120.81 & 117.25 & 120.01 & 120.47 & 118.86 & 118.84 & 118.83 & 119.22 & 122.62 \\
\texttt{TAE\_W4-17\_4} & 306.44 & 294.55 & 313.02 & 314.56 & 313.09 & 313.02 & 312.80 & 312.87 & 312.64 \\
\texttt{TAE\_W4-17\_5} & 163.02 & 159.89 & 158.79 & 159.69 & 157.70 & 157.69 & 157.69 & 158.17 & 163.79 \\
\texttt{TAE\_W4-17\_6} & 426.94 & 416.96 & 423.01 & 426.15 & 425.64 & 425.60 & 425.38 & 425.42 & 430.96 \\
\end{longtable}
\endgroup

\begingroup
\scriptsize
\setlength{\tabcolsep}{3.0pt}
\renewcommand{\arraystretch}{0.98}
\begin{longtable}{@{}llrrrrrrrrr@{}}
\caption{Complete MGBL20~\citep{zhao2008sogga,peverati2012xc} signed bond-length errors, in \AA{}, for the nine hydrogen-containing MGHBL9 bonds and 11 heavy-atom MGNHBL11 bonds; values are truncated to three decimal places before subtraction and the experimental lengths are in the final column.
Errors are calculated minus reference.
Aggregate ME and MAE values below are reported in m\AA{}.
Starred LAK values are from Ref.~\citenum{lebeda2024lak} and were not calculated here.
Random-Search Best is represented as RS Best.}\label{tab:focused-mgbl20}\\
\toprule
Molecule & Bond & PBE & SCAN & r2SCAN & r2SCAN+ & \shortstack{Finalist\\2} & \shortstack{Finalist\\3} & RS Best & LAK$^{*}$ & Reference (\AA) \\
\midrule
\endfirsthead
\multicolumn{11}{c}{\tablename\ \thetable\ (continued)}\\
\toprule
Molecule & Bond & PBE & SCAN & r2SCAN & r2SCAN+ & \shortstack{Finalist\\2} & \shortstack{Finalist\\3} & RS Best & LAK$^{*}$ & Reference (\AA) \\
\midrule
\endhead
\midrule
\multicolumn{11}{r}{Continued on next page}\\
\endfoot
\bottomrule
\endlastfoot
\multicolumn{11}{l}{\textit{MGHBL9 (9)}} \\
C$_{2}$H$_{2}$ & C--H & +0.007 & -0.002 & -0.001 & +0.000 & +0.000 & +0.000 & +0.000 & +0.000 & 1.063 \\
CH$_{4}$ & C--H & +0.009 & +0.000 & +0.001 & +0.001 & +0.001 & +0.001 & +0.001 & +0.003 & 1.086 \\
H$_{2}$ & H--H & +0.009 & +0.000 & +0.000 & +0.000 & +0.000 & +0.000 & +0.000 & +0.000 & 0.741 \\
H$_{2}$CO & C--H & +0.015 & +0.005 & +0.007 & +0.009 & +0.009 & +0.009 & +0.008 & +0.008 & 1.102 \\
H$_{2}$O & O--H & +0.011 & +0.002 & +0.002 & +0.002 & +0.002 & +0.002 & +0.002 & +0.003 & 0.957 \\
HCN & C--H & +0.009 & +0.001 & +0.002 & +0.003 & +0.003 & +0.003 & +0.002 & +0.002 & 1.065 \\
HF & H--F & +0.013 & +0.003 & +0.003 & +0.003 & +0.003 & +0.003 & +0.003 & +0.003 & 0.917 \\
\end{longtable}
\endgroup

\clearpage
\begingroup
\fontsize{7.4}{8.8}\selectfont
\setlength{\tabcolsep}{1.1pt}
\renewcommand{\arraystretch}{0.98}
\begin{longtable}{@{}lrrrrrrrrrrrr@{}}
\caption{Fundamental gaps, in eV, for the 24-solid union of TASK14~\citep{aschebrock2019ultranonlocality} (asterisks) and SCBG15~\citep{lebeda2024lak} (daggers); doubly marked systems occur in both.
The TASK, mTASK, and LAK values were not calculated here but were taken from the respective papers cited in the column headings; dashes indicate unavailable values.
TASK used the all-electron BAND code, mTASK used VASP with PAW datasets, and LAK used scalar-relativistic all-electron BAND, whereas our calculations used Quantum ESPRESSO with SG15 pseudopotentials.
The comparison therefore aligns systems and experimental references but is not a same-code or same-protocol ranking.
Experimental and HSE06~\citep{heyd2003hse,krukau2006hse06} values follow the revised MBPT benchmark~\citep{grossmann_mbpt,borlido2019bandgaps}, except that CdO and MgS use TASK experiments and HSE06 values from Refs.~\citenum{kim2020bandgapdatabase} and \citenum{lucero2012hiss}, respectively, while ZnTe follows Ref.~\citenum{lebeda2024lak}.
Random-Search Best is represented as RS Best.}\label{tab:focused-bandgap-union}\\
\toprule
Solid & PBE & SCAN & r2SCAN & r2SCAN+ & \shortstack{Finalist\\2} & \shortstack{Finalist\\3} & RS Best & TASK~\citep{aschebrock2019ultranonlocality} & mTASK~\citep{neupane2021mtask} & LAK~\citep{lebeda2024lak} & HSE06 & Expt. \\
\midrule
\endfirsthead
\multicolumn{13}{c}{\tablename\ \thetable\ (continued)}\\
\toprule
Solid & PBE & SCAN & r2SCAN & r2SCAN+ & \shortstack{Finalist\\2} & \shortstack{Finalist\\3} & RS Best & TASK~\citep{aschebrock2019ultranonlocality} & mTASK~\citep{neupane2021mtask} & LAK~\citep{lebeda2024lak} & HSE06 & Expt. \\
\midrule
\endhead
\midrule
\multicolumn{13}{r}{Continued on next page}\\
\endfoot
\bottomrule
\endlastfoot
Ge$^{*\dagger}$ & 0.08 & 0.12 & 0.00 & 0.00 & 0.00 & 0.00 & 0.00 & 0.85 & 1.00 & 0.83 & 0.86 & 0.74 \\
CdO$^*$ & 0.00 & 0.00 & 0.00 & 0.34 & 0.34 & 0.33 & 0.25 & 0.89 & 0.75 & -- & 0.76 & 0.84 \\
Si$^{*\dagger}$ & 0.56 & 0.83 & 0.70 & 0.72 & 0.71 & 0.72 & 0.72 & 1.12 & 1.46 & 1.15 & 1.20 & 1.17 \\
GaAs$^{*\dagger}$ & 0.52 & 0.45 & 0.60 & 0.76 & 0.75 & 0.75 & 0.68 & 1.72 & 2.07 & 1.57 & 1.46 & 1.52 \\
SiC$^{*\dagger}$ & 1.39 & 1.70 & 1.72 & 1.95 & 1.95 & 1.95 & 1.90 & 2.06 & 2.05 & 2.14 & 2.25 & 2.42 \\
ZnS$^{*\dagger}$ & 2.13 & 2.59 & 2.64 & 3.10 & 3.10 & 3.09 & 2.98 & 3.84 & 3.96 & 3.18 & 3.38 & 3.72 \\
MgS$^*$ & 3.55 & 4.16 & 4.18 & 4.70 & 4.70 & 4.70 & 4.62 & 5.74 & 6.23 & -- & 4.48 & 5.40 \\
C$^*$ & 4.17 & 4.48 & 4.25 & 4.26 & 4.26 & 4.27 & 4.28 & 4.36 & 4.98 & -- & 5.34 & 5.50 \\
\end{longtable}
\endgroup

\begingroup
\scriptsize
\setlength{\tabcolsep}{4.2pt}
\renewcommand{\arraystretch}{0.98}
\begin{longtable}{@{}lrrrrrrrr@{}}
\caption{Calculated equilibrium lattice constants and zero-point-vibrational-energy-corrected experimental references for the 20-solid LC20 set, in \AA.
Reference values are taken from the assessment of Csonka et al.~\citep{csonka2009}.
ME is calculated minus experiment; statistics use unrounded fitted values.
Random-Search Best is represented as RS Best.}\label{tab:focused-lc20}\\
\toprule
Solid & PBE & SCAN & r2SCAN & r2SCAN+ & \shortstack{Finalist\\2} & \shortstack{Finalist\\3} & RS Best & \shortstack{ZPVE-corrected\\reference} \\
\midrule
\endfirsthead
\multicolumn{9}{c}{\tablename\ \thetable\ (continued)}\\
\toprule
Solid & PBE & SCAN & r2SCAN & r2SCAN+ & \shortstack{Finalist\\2} & \shortstack{Finalist\\3} & RS Best & \shortstack{ZPVE-corrected\\reference} \\
\midrule
\endhead
\midrule
\multicolumn{9}{r}{Continued on next page}\\
\endfoot
\bottomrule
\endlastfoot
Li & 3.436 & 3.414 & 3.432 & 3.463 & 3.464 & 3.465 & 3.466 & 3.449 \\
Na & 4.200 & 4.200 & 4.228 & 4.316 & 4.317 & 4.318 & 4.314 & 4.210 \\
Ca & 5.524 & 5.550 & 5.569 & 5.679 & 5.680 & 5.679 & 5.671 & 5.553 \\
Sr & 6.025 & 6.077 & 6.090 & 6.264 & 6.265 & 6.264 & 6.248 & 6.045 \\
Ba & 5.019 & 5.036 & 5.061 & 5.244 & 5.244 & 5.239 & 5.205 & 4.995 \\
Al & 4.043 & 4.000 & 3.997 & 4.001 & 4.001 & 4.002 & 4.005 & 4.020 \\
C & 3.568 & 3.556 & 3.560 & 3.568 & 3.568 & 3.568 & 3.566 & 3.543 \\
Si & 5.478 & 5.447 & 5.459 & 5.485 & 5.485 & 5.485 & 5.481 & 5.416 \\
\end{longtable}
\endgroup

\begin{table}[H]
\centering
\scriptsize
\setlength{\tabcolsep}{4pt}
\renewcommand{\arraystretch}{1.1}
\caption{Benchmark summary for r2SCAN, r2SCAN+, and Random-Search Best.}
\label{tab:focused-manual-benchmark-summary}
\begin{tabular}{@{}lrrr@{}}
\toprule
Benchmark & r2SCAN & r2SCAN+ & Random-Search Best \\
\midrule
SCBG15 band-gap MAE (eV) & 0.63 & 0.39 & 0.45 \\
TASK14 band-gap MAE (eV) & 1.78 & 1.43 & 1.50 \\
24-solid band-gap MAE (eV) & 1.26 & 0.96 & 1.02 \\
MGBL20 bond-length MAE (m\AA{}) & 3.7 & 3.3 & 3.4 \\
LC20 lattice-constant MAE (\AA{}) & 0.029 & 0.075 & 0.069 \\
All 329 molecular reactions MAE (kcal/mol) & 4.79 & 4.42 & 4.37 \\
\bottomrule
\end{tabular}
\end{table}

\nocite{kim2020bandgapdatabase,lucero2012hiss}
\masterbibcontrols
\begingroup
\makeatletter
\let\hyper@natanchorstart\@gobble
\let\hyper@natanchorend\relax
\let\label\@gobble
\makeatother
\section*{References}
\putbib
\endgroup
\end{bibunit}

\end{document}